\documentclass[letterpaper]{aa}
\usepackage{graphicx}
\usepackage{txfonts}
\usepackage{natbib}
\usepackage{amssymb}
\newcommand{\lya}{Ly$\alpha$}
\newcommand{\ergscm}{erg\,s$^{-1}$\,cm$^{-2}$}
\newcommand{\kms}{km\,s$^{-1}$}
\newcommand{\msunyr}{{\em M}$_{\sun}$\,yr$^{-1}$}
\newcommand{\msunyrmpc}{{\em M}$_{\sun}$\,yr$^{-1}$\,Mpc$^{-3}$}

\begin{document}
   \title{Properties of \lya\ emitters around the radio galaxy MRC
0316--257\thanks{Based on observations carried out at the European
Southern Observatory, Paranal, Chile, programs LP167.A-0409 and
68.B-0295.}}


   \author{B.\ P.\ Venemans \inst{1} \and H.\ J.\ A.\ R\"ottgering
     \inst{1} \and G.\ K.\ Miley \inst{1} \and J.\ D.\ Kurk \inst{2}
     \and C.\ De Breuck \inst{3} \and R.\ A.\ Overzier \inst{1} \and
     W.\ J.\ M.\ van Breugel \inst{4} \and C.\ L.\ Carilli \inst{5} \and
     H.\ Ford \inst{6} \and T.\ Heckman \inst{6} \and L.\ Pentericci
     \inst{7} \and P.\ McCarthy \inst{8}}

   \offprints{B.\ P.\ Venemans (venemans@strw.leidenuniv.nl)}

   \institute{Sterrewacht Leiden, P.O. Box 9513, 2300 RA, Leiden, 
              The Netherlands               
         \and
              INAF, Osservatorio Astrofisico di Arcetri, Largo Enrico
              Fermi 5, 50125, Firenze, Italy
	 \and
              European Southern Observatory, Karl Schwarzschild
              Stra{\ss}e 2, D-85748 Garching, Germany
         \and
              Lawrence Livermore National Laboratory, 
              P.O. Box 808, Livermore CA, 94550, USA
         \and
              NRAO, P.O. Box 0, Socorro NM, 87801, USA
         \and
              Dept. of Physics \& Astronomy, The Johns Hopkins University,
              3400 North Charles Street, Baltimore MD, 21218-2686, USA
         \and
              Dipartimento di Fisica, Universit\`a degli studi Roma Tre,
              Via della Vasca Navale 84, Roma, 00146, Italy
         \and
              The Observatories of the Carnegie Institution of Washington,
              813 Santa Barbara Street, Pasadena CA, 91101, USA
             }

   \date{Received 21 September 2004; accepted 21 October 2004}

   \authorrunning{B.\ P.\ Venemans et al.}
   \titlerunning{Properties of \lya\ emitters around MRC 0316--257}

   \abstract{Observations of the radio galaxy MRC 0316--257 at $z =
   3.13$ and the surrounding field are presented. Using narrow- and
   broad-band imaging obtained with the VLT, 77 candidate
   \lya\ emitters with a rest-frame equivalent width of $>$ 15 \AA\
   were selected in a $\sim 7$\arcmin$\times7$\arcmin\ field around
   the radio galaxy. Spectroscopy of 40 candidate emitters resulted in
   the discovery of 33 emission line galaxies of which 31 are \lya\
   emitters with redshifts similar to that of the radio galaxy, while
   the remaining two galaxies turned out to be [\ion{O}{ii}]
   emitters. The \lya\ profiles have widths ({\em FWHM}) in the range of
   120--800 \kms, with a median of 260 \kms. Where the signal-to-noise
   was large enough, the \lya\ profiles were found to be asymmetric,
   with apparent absorption troughs blueward of the profile peaks,
   indicative of absorption along the line of sight of an \ion{H}{i}
   mass of at least $2 \times 10^2 - 5 \times 10^4$ {\em M}$_{\sun}$.
   Besides that of the radio galaxy and one of the emitters that is a
   QSO, the continuum of the emitters is faint, with luminosities
   ranging from 1.3 $L^*$ to $<$ 0.03 $L^*$. The colors of the
   confirmed emitters are, on average, very blue. The median UV
   continuum slope is $\beta = -1.76$, bluer than the average slope of
   LBGs with \lya\ emission ($\beta \sim -1.09$). A large fraction of
   the confirmed emitters ($\sim 2/3$) have colors consistent with
   that of dust-free starburst galaxies. Observations with the
   Advanced Camera for Surveys on the {\em Hubble Space Telescope}
   show that the emitters that were detected in the ACS image have a
   range of different morphologies. Four \lya\ emitters ($\sim 25$\%)
   were unresolved with upper limits on their half light radii of $r_h
   < 0.6-1.3$ kpc, three objects ($\sim 19$\%) show multiple clumps of
   emission, as does the radio galaxy, and the rest ($\sim 56$\%) are
   single, resolved objects with $r_h < 1.5$ kpc. A comparison with
   the sizes of Lyman break galaxies at $z \sim 3$ suggests that the
   \lya\ emitters are on average smaller than LBGs. The average star
   formation rate of the \lya\ emitters is 2.6 \msunyr\ as measured by
   the \lya\ emission line or $<$ 3.9 \msunyr\ as measured by the UV
   continuum. The properties of the \lya\ galaxies (faint, blue and
   small) are consistent with young star forming galaxies which are
   still nearly dust free.

   The volume density of \lya\ emitting galaxies in the field around MRC
   0316--257 is a factor of $3.3^{+0.5}_{-0.4}$ larger compared with
   the density of field \lya\ emitters at that redshift. The velocity
   distribution of the spectroscopically confirmed emitters has a
   dispersion of 640 \kms, corresponding to a FWHM of 1510 \kms, which
   is substantially smaller than the width of the narrow-band filter
   (FWHM $\sim 3500$ \kms). The peak of the velocity distribution is
   located within 200 \kms\ of the redshift of the radio galaxy. We
   conclude that the confirmed \lya\ emitters are members of a
   protocluster of galaxies at $z \sim 3.13$. The size of the
   protocluster is unconstrained and is larger than $3.3 \times 3.3$
   Mpc$^2$. The mass of this structure is estimated to be $>$ 3--6
   $\times 10^{14}$ {\em M}$_{\sun}$ and could be the progenitor of a cluster
   of galaxies similar to e.g.\ the Virgo cluster.

   }

   \maketitle
%

\section{Introduction}

Within Cold Dark Matter (CDM) scenarios the first stars and stellar
systems form through gravitational infall of primordial gas in large
CDM halos \citep[e.g.][]{whi78}. Numerical simulations suggest that as
these halos merge they form vast, web-like networks of young galaxies
and ionized gas \citep[e.g.][]{bau98}. The most massive galaxies, and
the richest clusters emerge from regions with the largest
overdensities. Although clusters of galaxies have been studied
extensively out to $z \sim 1.3$
\citep[e.g.][]{ros99,del00,sta02,bla03a,mau03,tof04}, the epoch of
cluster formation is still an open question due to the difficulty in
identifying their progenitors in the early Universe. 

During the last decade, evidence has mounted that the most powerful
high redshift radio galaxies (HzRGs; $z > 2$) are progenitors of
brightest cluster galaxies and are located in dense environments.
HzRGs are amongst the brightest and presumably most massive galaxies
\citep{jar01,deb02,zir03}. They have high star formation rates ($>
100$ \msunyr), based on deep spectra of their UV continuum
\citep[e.g.][]{dey97} and the detections of dust
\citep[e.g.][]{arc01,stev03,reu04} and extended CO emission
\citep{pap00,deb03a,deb03b}. Furthermore, radio galaxies at redshifts
between 0.5 and 1.5 are known to predominantly lie in moderately rich
clusters \citep{hil91,bes00,bes03}. At higher redshifts ($z > 2$),
some radio galaxies were found to posses companion galaxies
\citep{lef96,pas96,rot96,kee99}. Also, 20\% of the HzRGs have extreme
radio rotation measures ($>$ 1000 rad\,m$^{-2}$), giving an indication
that these radio galaxies are surrounded by dense hot gas
\citep{car97,ath98,pen00b}.

To search for direct evidence of the association of a cluster or a
forming cluster (protocluster) with a radio galaxy, we conducted a
pilot project on the Very Large Telescope (VLT) aimed at finding an
excess of \lya\ emitters around the clumpy radio galaxy PKS 1138--262
at $z = 2.16$. Narrow-band imaging resulted in a list of $\sim 50$
candidate \lya\ emitters \citep{kur00,kur04}. Subsequent multi-object
spectroscopy confirmed 14 \lya\ emitting galaxies and one QSO whose
velocities were within 1000 \kms\ of the central radio galaxy
\citep{pen00a,kur04}. The volume density of \lya\ emitters near PKS
1138--262 was found to be a factor $4.4 \pm 1.2$ times that of \lya\
emitters in blank fields \citep{kur04}. Using near-infrared narrow-
and broad-band images of the field, significant populations of
H$\alpha$ emitters at the redshift of the radio galaxy and extremely
red objects were found. Also, {\em Chandra} observations revealed an
excess of soft X-ray sources in the field of PKS 1138--262
\citep{pen02}, indicating that several AGN are present in the
protocluster.

As shown by the study of the overdense region near PKS 1138--262,
distant protoclusters provide ideal laboratories for tracing the
development of large scale structure and galaxy evolution. To further
study the formation of large scale structure in the early Universe and
to investigate the evolution of galaxies in dense environments, we
initiated a large program on the VLT to search for \lya\ emitting
galaxies around luminous radio galaxies with redshifts $2 < z < 5$
\citep{ven03}. The goals were to find protoclusters of galaxies,
determine the fraction of HzRGs associated with protoclusters and
study the properties of protoclusters and their galaxies. The first
result was the discovery of a protocluster around the radio galaxy TN
J1338--1942 at $z = 4.1$ \citep{ven02}. Deep imaging and spectroscopy
revealed 20 \lya\ emitters within a projected distance of 1.3 Mpc and
600 \kms\ of the radio galaxy. By comparing the density of \lya\
emitters in the protocluster to the field, the galaxy overdensity was
claimed to be $4.0 \pm 1.4$ and the mass of the structure was
estimated to be $\sim 10^{15}$ {\em M}$_{\sun}$ \citep{ven02}.

Here we report on observations of the radio galaxy MRC 0316--257. This
1.5 Jy radio source was listed in the 408 MHz Molonglo
Reference Catalogue \citep{lar81} and optically identified by
\citet{mcc90}. Its discovery spectrum yielded a redshift of 3.13
\citep{mcc90}. This object was included in our program because the
redshift of the radio galaxy shifted the \lya\ line into one of the
narrow-band imaging filters available at the VLT. Also, it already had
two spectroscopically confirmed \lya\ emitting companions
\citep[][hereafter LF96]{lef96}, an indication that the radio galaxy
is located in a dense environment. Further, the redshift of the radio
galaxy of 3.13 allows for an efficient search for Lyman Break Galaxies
(LBGs) and for [\ion{O}{iii}] $\lambda 5007$ \AA\ emitters using a
$K$-band narrow-band filter, which is available in the Infrared
Spectrometer and Array Camera \citep[ISAAC,][]{moo97} at the VLT. 

Besides observing MRC 0316--257 with the VLT as part of our large
program, we made additional observations of the field with the
Advanced Camera for Surveys \citep[ACS;][]{for98} on the {\em Hubble
Space Telescope} {\em (HST)}\footnote{Based on observations made with
the NASA/ESA Hubble Space Telescope, obtained at the Space Telescope
Science Institute, which is operated by the Association of
Universities for Research in Astronomy, Inc., under NASA contract NAS
5-26555. These observations are associated with program \#8183} to
study the sizes and morphologies of the detected galaxies.

This paper is organized in the following way. In Sect.\ \ref{obs} the
imaging observations and data reduction are described and Sect.\
\ref{select} discusses how candidate \lya\ emitters in the field are
detected. The spectroscopic observations and the results are presented
in Sect.\ \ref{spec}. The properties of the \lya\ emitters are
analyzed in Sect.\ \ref{properties}, and details of individual
emitters are presented in Sect.\ \ref{notes}. Evidence for the
presence of a protocluster in the field is discussed in Sect.\
\ref{protocluster}, and the properties are presented in Sect.\
\ref{protoprop}. In Sect.\ \ref{nature} the nature of the \lya\
emitters is discussed, followed by a description of the implications of a
protocluster at $z=3.13$ in Sect.\ \ref{implications}.

Throughout this article, magnitudes are in the AB system
\citep{oke74}, using the transformations $V_{\mathrm{AB}} =
V_{\mathrm{Vega}} +0.01$ and $I_{\mathrm{AB}} = I_{\mathrm{Vega}}
+0.39$ \citep{bes79}. A $\Lambda$-dominated cosmology with H$_0 = 65$
\kms\ Mpc$^{-1}$, $\Omega_\mathrm{M} = 0.3$, and $\Omega_{\Lambda} =
0.7$ is assumed. In this cosmology, the luminosity distance of MRC
0316--257 is $28.8$ Gpc and 1\arcsec\ corresponds to 8.19
kpc at $z=3.13$.

\section{Imaging observations and data reduction}
\label{obs}

\begin{table*}
\begin{center}
\caption{Summary of the observations of the field around MRC 0316--257.}
\begin{tabular}{lllllll}
\hline
\hline
Date & Telescope & Instrument & Mode & Optical element & Seeing &
Exposure time \\
\hline
2001, September 20 and 21 & VLT UT4 & FORS2 & Imaging & Bessel $V$ &
0\farcs7 & 4\,860 s \\
2001, September 20 and 21 & VLT UT4 & FORS2 & Imaging & OIII/3000 &
0\farcs7 & 23\,400 s \\
2001, September 22 & VLT UT4 & FORS2 & MOS$^a$ & GRIS\_1400V & $\sim$
1\farcs5 & 12\,600 s \\
2001, October 18 & VLT UT4 & FORS2 & MXU$^b$, mask I & GRIS\_1400V &
1\farcs0 & 10\,800 s \\
2001, October 18, 19 and 20 & VLT UT4 & FORS2 & MXU$^b$, mask II &
GRIS\_1400V & 1\farcs0 & 29\,100 s \\
2001, November 15 and 16 & VLT UT3 & FORS1 & PMOS$^c$ & GRIS\_300V &
0\farcs8 & 19\,800 s \\
2002, July 18 & {\em HST} & ACS & Imaging & F814W & -- & 6300 s \\ 
2002, September 6, 7 and 8 & VLT UT4 & FORS2 & Imaging & Bessel $I$ &
0\farcs7 & 4\,680 s \\
\hline
\end{tabular}
\end{center}
$^a$ Multi-object spectroscopy mode, performed with 19 movable slitlets
with lengths of 20\arcsec--22\arcsec. \\
$^b$ Multi-object spectroscopy mode with a user-prepared mask.\\ 
$^c$ Spectropolarimetry mode using 9 movable slitlets of 20\arcsec.
\end{table*}

\subsection{VLT imaging}
\label{ima}

An overview of the observations is shown in Table 1. On
2001, September 20 and 21, narrow- and broad-band imaging was carried
out with the 8.2 m Yepun (VLT UT4) to search for \lya\ emitting
galaxies around MRC 0316--257. The instrument used was the FOcal
Reducer/low dispersion Spectrograph 2 \citep[FORS2;][]{app92} in
imaging mode. For the narrow-band imaging the OIII/3000 filter was
used with a central wavelength of 5045 \AA\ and full width half
maximum ({\em FWHM}) of 59 \AA, which samples the \lya\ line from the radio
galaxy, which is redshifted to 5021 \AA\ \citep[][LF96]{mcc90}. To
measure the UV continuum near the \lya\ line, the field was imaged
with broad-band filter Bessel $V$ with a central wavelength
of 5540 \AA\ and {\em FWHM} of 1115 \AA. The detector was a SiTE CCD with
2048$\times$2048 pixels. The pixel scale was 0\farcs2 per pixel,
resulting in a field of view of 6\farcm8$\times$6\farcm8. A year
later, on 2002, September 6, 7 and 8, broad-band images of the field
were taken in the Bessel $I$ filter, with a central wavelength of 7680
\AA\ and a {\em FWHM} of 1380 \AA. The instrument was again FORS2, but the
detector was replaced by two MIT CCDs with 2048$\times$2048 pixels
each. The gap between the two CCDs was $\sim 4$ arcsec. The pixel
scale of the MIT CCDs was 0\farcs125 per pixel. To decrease the
readout time, the pixels were binned by $2\times2$, resulting in a
spatial scale of 0\farcs25 pixel$^{-1}$. The field of view was
restricted by the geometry of the Multi-Object Spectroscopy (MOS)
unit, and was 6\farcm8$\times$6\farcm8.

The observations in the narrow-band were split into 13 separate
exposures of 1800 s, in $V$-band into 27 separate 180 s exposures and
in the $I$-band 26 exposures of 180 s were taken. The individual
exposures were shifted by $\sim 15$\arcsec\ with respect to each other
to facilitate identifying cosmic rays and removing residual flat-field errors.

All nights except for 2001, September 20 were photometric, and the
average seeing was 0\farcs65 - 0\farcs7 in the narrow-band, $V$ and
$I$ images (see Table 1).  For the flux calibration, the
spectrophotometric standard star LTT 1788 \citep{sto83,bal84} was
observed in the $V$-band and the photometric standard stars in the
field SA98 \citep{lan92} were used to calibrate the $I$-band images.

\subsection{Data reduction of VLT data}
\label{vltdatared}

The VLT images were reduced using standard routines within the
reduction software package IRAF\footnote{IRAF is distributed by the
National Optical Astronomy Observatories, which are operated by the
Association of Universities for Research in Astronomy, Inc., under
cooperative agreement with the National Science Foundation.}. The
reduction steps included bias subtraction, flat fielding using
twilight sky flats and illumination correction using the unregistered
science frames.

The magnitude zero-points derived from different standard stars were
consistent with each other within 0.02 magnitude. To derive the
zero-point of the narrow-band image, the magnitude of the $\sim 400$
brightest objects in the field were measured in the $V$ and $I$-band
images. These objects had a signal-to-noise of at least 25 in both $V$
and $I$-band images. A magnitude limit of $m_I > 20$ was set to reject
saturated stars. Narrow-band magnitudes were derived from the $V$ and
$I$-band magnitudes assuming a powerlaw spectral energy distribution
for these 400 objects. With these derived narrow-band magnitudes and
the associated counts in the narrow-band image, the zero-point of the
narrow-band image was computed. The rms of the computed zero-point was
0.006 mag.

All magnitudes were corrected for galactic extinction which was
estimated by \citet{sch98} to have a value of $E(B-V)=0.014$ mag. The
measured 1 $\sigma$ limiting magnitudes per square arcsecond were
28.35, 28.90 and 28.69 for the narrow-band, $V$-band and $I$-band
respectively.

Astrometric calibration was performed using the USNO-A2.0 catalog
\citep{mon98a,mon98b} from which 20 stars were identified in the
field. This resulted in a fit with a typical error in the right
ascension and declination of 0\farcs17. The astrometric accuracy of
the images is dominated by the uncertainty of the USNO-A2.0 catalog
of 0\farcs25 \citep{deu99}. The VLT images were registered in the
following way. Because the $V$-band and narrow-band images were taken
with the same detector, a simple pixel shift was sufficient to align
the images, using the positions of a few stars over the field. The
$I$-band frames were taken with the MIT CCDs, which had a different
pixel scale (0\farcs25 pixel$^{-1}$) as compared to the SiTE CCD
(0\farcs2 pixel$^{-1}$). The distortions were also different, and
rescaling the $I$-band image to the same pixel scale as the $V$-band
image resulted in positional errors up to a few arcseconds in the
corners of the $I$-band image. Instead, the position of a few hundred
objects detected in both the $I$-band and $V$-band images with a
signal-to-noise greater than 15, were used for the alignment. This way
the positional error of objects in the $I$-band image dropped to
0\farcs04. Subsequently, the algorithm DRIZZLE \citep{fru02} was used
to map the $I$, $V$ and narrow-band images on new frames with a common
pixel scale of 0\farcs16.

The area of the reduced images was 46.7 arcmin$^2$. Due to the
presence of two bright stars in the field, the area that could be used
for detecting candidate \lya\ emitters was 45.75 arcmin$^2$. The width
of the narrow-band filter in redshift is 0.049 at $z \sim 3.13$ and the
volume probed by the filter at $z=3.13$ is 9331 Mpc$^3$.

\subsection{Hubble Space Telescope imaging and reduction}

A part of the field imaged by the VLT was observed in 2002 July with
the ACS on board the {\em HST} as part of an imaging program of
HzRGs. The 3\farcm4$\times$3\farcm4 field of view of the ACS was
chosen to include not only the radio galaxy but also as many confirmed
\lya\ emitters as possible (see Fig.\ \ref{skydist} for the position
of the ACS field within the FORS field). The field was imaged in the
F814W filter (hereafter $I_{814}$) with a central wavelength of 8333
\AA\ and a width of 2511 \AA. The total exposure time was 6300 s. The
images were reduced using the ACS GTO pipeline \citep{bla03b}.

\section{Detection and selection of candidate emitters}
\label{select}

\subsection{Source detection}

\begin{figure}
\resizebox{\hsize}{!}{\includegraphics{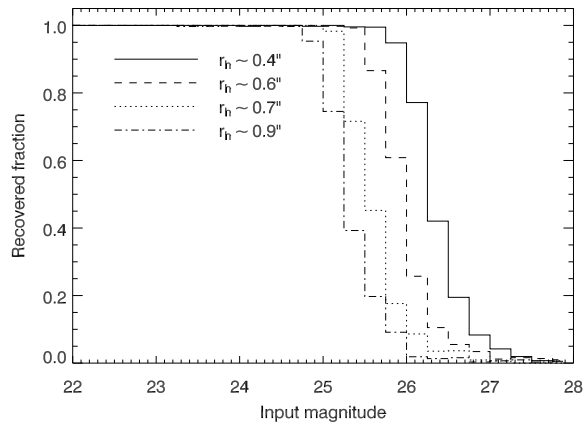}}
\caption{Fraction of galaxies recovered in the narrow-band image
  versus magnitude for various galaxy sizes.}
\label{completeness}
\end{figure}

For the detection and photometry of objects in the images, the program
SExtractor \citep[version 2.2.2,][]{ber96} was applied. The
narrow-band image was used to detect the objects. Because some of the
\lya\ emitters remained undetected in the broad-band images (see Table
4), this was preferred above a combination of the
narrow-band and broad-band images as detection image, which is favoured
by some other groups \citep[e.g.][]{fyn02}. Detected objects in the
narrow-band image were defined to have at least 15 connected pixels
with values larger than the rms sky noise. This resulted in a list of
3505 objects, of which 3209 had a signal-to-noise greater than
five. To assess the completeness of the catalog, artificial and real
galaxies were added to the narrow-band image and recovered. The
galaxies had various sizes, the smallest galaxies had a half light
radius $r_h \sim 0.4$\arcsec, similar to that of stars in the field
(unresolved objects), the largest galaxies had a half light radius
$r_h \sim 0.9$\arcsec, roughly 2.5 times that of stars. We found that
the completeness depended on the size of the galaxies that were added
to the image as shown in Fig.\ \ref{completeness}. The limit where
half of the galaxies were recovered ranged from a magnitude of
$m_\mathrm{nb} \lesssim 26.25$ for unresolved objects to
$m_\mathrm{nb} \lesssim 25.25$ for the largest objects. The 90\%
recovery limit was $m_\mathrm{nb} \sim 26.0$ for unresolved and
$m_\mathrm{nb} \sim 25.0$ for the largest objects.

\subsection{Photometry}
\label{photometry}

To measure the flux of the detected objects, the double image mode of
SExtractor was employed. In this mode, SExtractor detects objects in
one image, and carries out the photometry on a second image. In our
case, the narrow-band image was used for the detection of the objects
and the photometry was done on the narrow-band and broad-band images.
Of each detected object in the narrow-band the flux was measured in
two apertures: a circular aperture to compute the colors of the
object, and an elliptical aperture to estimate the total brightness of
the object. The radius of the circular aperture ($R_\mathrm{aper}$)
depended on the isophotal area of the object ($A_\mathrm{iso}$), which
is the area of pixels with values above the rms sky noise and is an
output parameter of SExtractor: $R_\mathrm{aper} =
\sqrt{A_\mathrm{iso}/\pi}$. A minimum aperture radius of 0\farcs525,
1.5 times the radius of the seeing disc, was set to avoid very small
apertures. The maximum radius was set to 4 times the radius of the
seeing disc to avoid overlapping apertures due to neighbouring
galaxies. The shape and size of the elliptical aperture was derived
from the object's light distribution. The ellipticity $\varepsilon$
and position angle of the object were computed from the second order
moment of the light distribution. Using the first moment $r_1$, the
elliptical aperture had major and minor axes of $k r_1 / \varepsilon$
and $\varepsilon k r_1$ \citep{ber96}. The scaling factor $k$
determines the size of aperture and is a free parameter in SExtractor.
We tested SExtractor on a set of images and compared SExtractor's
output magnitudes for values of $k$ in the range $1.0 < k < 2.75$. It
was found that a scaling factor of $k \approx 1.75$ both optimized the
signal-to-noise and minimized the fraction of the flux of the object
outside the aperture.

The aperture used to measure the total flux of an object was the
elliptical aperture, except when more than 10\% of the pixels in the
elliptical aperture was significantly effected by bright and close
neighbours (SExtractor output parameter FLAGS equals 1) or when the
object was originally blended with another one (FLAGS equals 2). In
those cases the circular aperture was used to derive the total flux.

\begin{figure}
\resizebox{\hsize}{!}{\includegraphics{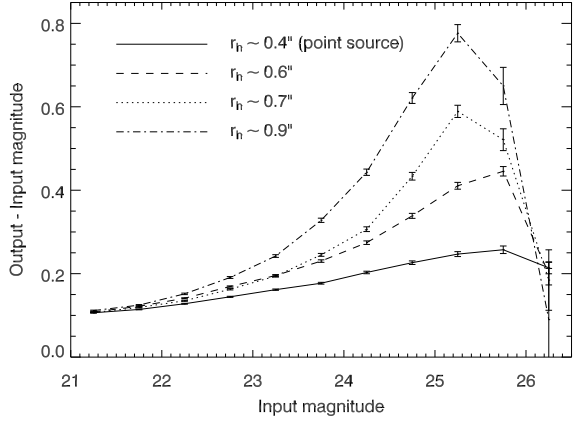}}
\caption{Difference between measured magnitude and input magnitude as
  a function of input magnitude for objects with various sizes. The
  difference is larger for fainter and/or larger objects.}
\label{correction}
\end{figure}

To estimate the fraction of the total flux of the object falling
outside the (elliptical) aperture, Monte Carlo simulations were
performed. Galaxies in the magnitude range $21 < m_\mathrm{nb} < 28$
with various shapes (Gaussian and elliptical profiles) and sizes (half
light radii 0.4\arcsec\ $< r_h <$ 0.9\arcsec) were added to the
narrow-band image and recovered. The number of galaxies added to the
image was limited to 40 to avoid overcrowding. This routine was
repeated until 8000 galaxies per magnitude bin were simulated. By
comparing the measured magnitudes to the input magnitude of the
simulated galaxies, the fraction of the flux that was outside the
aperture could be estimated to correct the aperture magnitudes. The
simulations showed that this fraction depends on both the original
size of object and the magnitude of the object: the fraction of the
flux outside the aperture is higher for a large and/or faint object,
compared to a compact and/or bright object (see Fig.\
\ref{correction}). At faint magnitudes the correction becomes smaller
again, because these objects have a larger probability to be detected
if they coincide with a peak in the noise. For the brightest objects
the fraction of the flux outside the aperture is constant at a value
of $\sim 11$\%. To avoid an overestimation of the magnitude
correction, we decided to use only point sources to measure the
correction (see Fig.\ \ref{correction}). The magnitude correction
applied to the sources in our field was $\lesssim 0.1$ for objects
with $m_\mathrm{nb} \lesssim 21$, rising to $\sim 0.25$ for objects
with $m_\mathrm{nb} \sim 26$.

\subsection{Selection of candidate \lya\ emitters}
\label{selection}

An efficient method of detecting \lya\ emitting galaxies at high
redshift is to select objects with a large line equivalent width
\citep[e.g.][]{cow98} using the narrow- and broad-band photometry. The
observed equivalent width {\em EW}$_{\mathrm{obs}}$ of a \lya\ line is
defined as:

\begin{equation}
\label{ewobssimp}
EW_{\mathrm{obs}} = F_\mathrm{Ly\alpha} / C_\mathrm{Ly\alpha}
\end{equation}

\noindent 
with $F_\mathrm{Ly\alpha}$ the flux of the \lya\ line and
$C_\mathrm{Ly\alpha}$ the UV continuum at the wavelength of the \lya\
line. Assume that a \lya\ line is observed in both a narrow-band
filter and a broad-band filter, then the flux density in the
narrow-band ($f_{\lambda,\mathrm{nb}}$) and broad-band
($f_{\lambda,\mathrm{bb}}$) can be described as:

\begin{eqnarray}
\label{fnbsimp}
f_{\lambda,\mathrm{nb}} &=& C(\lambda = \lambda_{\mathrm{eff,nb}}) +
F_\mathrm{Ly\alpha} / \Delta \lambda_{\mathrm{nb}} \\
\label{fbbsimp}
f_{\lambda,\mathrm{bb}} &=& C(\lambda = \lambda_{\mathrm{eff,bb}}) +
F_\mathrm{Ly\alpha} / \Delta \lambda_{\mathrm{bb}},
\end{eqnarray}

\noindent
where $C$ is the UV continuum and $\Delta \lambda_{\mathrm{bb(nb)}}$
is the width of broad-band (narrow-band) filter (Eq.\ \ref{dlam}) and
$\lambda_{\mathrm{eff,bb(nb)}}$ the effective wavelength of broad-band
(narrow-band) filter (Eq.\ \ref{leff}). 

The effective wavelength of a filter with transmission curve
$T(\lambda)$ is given by

\begin{equation}
\label{leff}
\lambda_\mathrm{eff} = \frac{\int~ \lambda\, T(\lambda)\, d\lambda}{\int~
T(\lambda)\,d\lambda}
\end{equation}

\noindent
and the width of the filter $\Delta \lambda$ by

\begin{equation}
\label{dlam}
\Delta \lambda = \int~T(\lambda)\,d\lambda / T_\mathrm{max},
\end{equation}

\noindent
with $T_\mathrm{max}$ the peak transmission of the filter. For a
top-hat filter, the effective wavelength is equal to the central
wavelength, the width equals the {\em FWHM}. 

If the central wavelengths of the narrow-band and broad-band filters
are roughly equal and the \lya\ line falls in the centre of the
filters, then eliminating either $C$ or $F_\mathrm{Ly\alpha}$ by
substituting Eq.\ (\ref{fnbsimp}) in Eq.\ (\ref{fbbsimp}) gives:

\begin{eqnarray}
F_\mathrm{Ly\alpha} &=& \frac{\Delta \lambda_{\mathrm{bb}} \Delta
  \lambda_{\mathrm{nb}} (f_{\lambda,\mathrm{nb}} -
  f_{\lambda,\mathrm{bb}})}{\Delta \lambda_{\mathrm{bb}} - \Delta
  \lambda_{\mathrm{nb}}} \\
C_\mathrm{Ly\alpha} &=& \frac{\Delta \lambda_{\mathrm{bb}}
  f_{\lambda,\mathrm{bb}} - \Delta \lambda_{\mathrm{nb}}
  f_{\lambda,\mathrm{nb}}}{\Delta \lambda_{\mathrm{bb}} - \Delta
  \lambda_{\mathrm{nb}}},
\end{eqnarray}

\noindent 
and using Eq.\ (\ref{ewobssimp}) results in an expression for
{\em EW}$_{\mathrm{obs}}$:

\begin{equation}
\label{ewbun}
EW_{\mathrm{obs}} = \frac{\Delta \lambda_{\mathrm{bb}} \Delta
  \lambda_{\mathrm{nb}} (f_{\lambda,\mathrm{nb}} -
  f_{\lambda,\mathrm{bb}})}{\Delta \lambda_{\mathrm{bb}}
  f_{\lambda,\mathrm{bb}} - \Delta \lambda_{\mathrm{nb}}
  f_{\lambda,\mathrm{nb}}}
\end{equation}

\noindent
\citep[e.g.][]{bun95,mal02}. Alternatively, Eq.\
(\ref{ewbun}) can be used if it is expected that the fraction
of the continuum flux falling in the filters that is absorbed by
foreground \ion{H}{i} is comparable to the fraction of the \lya\ line
that is extinguished by intergalactic absorption \citep[as assumed by
e.g.][]{mal02}.

If the central wavelengths of the narrow-band and broad-band filters
differ, as is the case with our filters, then the slope of the UV
continuum is needed to extrapolate the continuum strength from the
central wavelength of the broad-band to the central wavelength of the
narrow-band. Including an extra broad-band contribution redward of the
\lya\ line, the continuum slope can be calculated as well. Below is
described how the equivalent width of a $z = 3.13$ \lya\ emitter can
be computed using our available photometry.

Assume that a \lya\ emitter has a spectral energy distribution that
consists of a \lya\ line with flux $F_\mathrm{Ly\alpha}$ and a UV
continuum redward of the \lya\ line with strength $C$ and powerlaw
slope $\beta$ ($f_\lambda \propto \lambda^\beta$). The flux density in
the narrow-band ($f_{\lambda,\mathrm{nb}}$), $V$-band
($f_{\lambda,\mathrm{V}}$) and $I$-band ($f_{\lambda,\mathrm{I}}$) can
then be characterized as:

\begin{eqnarray}
\label{fnb}
f_{\lambda,\mathrm{nb}} &=&
Q_\mathrm{nb}\,C\,\lambda_\mathrm{eff,nb}^\beta~+~\epsilon_\mathrm{nb}\,
F_{\mathrm{Ly}\alpha} / \Delta \lambda_\mathrm{nb} \\ 
\label{fv}
f_{\lambda,\mathrm{V}} &=&
Q_\mathrm{V}\,C\,\lambda_\mathrm{eff,V}^\beta~+~\epsilon_\mathrm{V}\,
F_{\mathrm{Ly}\alpha} / \Delta \lambda_\mathrm{V} \\ 
\label{fi}
f_{\lambda,\mathrm{I}} &=& C\,\lambda_\mathrm{eff,I}^\beta
\end{eqnarray}

\noindent
with $\lambda_\mathrm{eff,nb/V/I}$ the effective wavelength corresponding to
the narrow-band, $V$ and $I$ filter respectively (Eq.\ \ref{leff}),
$\Delta \lambda$ the width of the filter (Eq.\ \ref{dlam}), $\epsilon$
the efficiency of the filter at the wavelength of the redshifted \lya\
line and $Q$ the fraction of the continuum flux falling in the filter
that is absorbed by the \lya\ forest (Eq.\ \ref{Qfilter}). It should
be mentioned that, in contrast to Eq.\ (\ref{ewbun}), no correction
factor for foreground absorption of the \lya\ line is applied in this
calculation. If foreground extinction of the \lya\ line is taken into
account, then the equivalent width and \lya\ line flux will be higher
by $\sim 60$\% (see Eq.\ \ref{tau}).

For the filters (and instrument) used in this project, the input parameters
are: $\lambda_\mathrm{eff,nb} = 5040.1$ \AA, $\Delta
\lambda_\mathrm{nb} = 61.1$ \AA, $\lambda_\mathrm{eff,V} = 5561.9$
\AA, $\Delta \lambda_\mathrm{V} = 1145.6$ \AA\ and
$\lambda_\mathrm{eff,I} = 7946.5$ \AA. The efficiency $\epsilon$ of
the filters depends on the redshift of the \lya\ line. For all objects
a redshift of $z = 3.13$ is assumed, and the efficiencies are
$\epsilon_\mathrm{nb} = 0.76$ and $\epsilon_\mathrm{V} = 0.74$. It
should be stressed that the computed equivalent width does not depend
strongly on the assumed redshift in the interval $z= 3.13 -
3.17$. Assuming a redshift of $z = 3.12$ will yield equivalent widths
that are a factor of $\sim 2$ higher compared to the equivalent widths
computed with $z = 3.13$.

The fraction of the continuum flux that is absorbed by foreground
neutral hydrogen averaged over the bandpass is $Q$:

\begin{equation}
Q = \frac{\int e^{-\tau_\mathrm{eff}} T(\lambda) \mathrm{d}\lambda}{\int
T(\lambda) \mathrm{d}\lambda}
\label{Qfilter}
\end{equation}

\noindent
where $\tau_\mathrm{eff}$ is the effective opacity of \ion{H}{i}. For
observed wavelengths between the redshifted \lya\ and redshifted
Ly$\beta$ line ($\lambda_{\mathrm{Ly\beta}} (1 + z) <
\lambda_\mathrm{obs} < \lambda_\mathrm{Ly\alpha} (1+z)$), the
expression for $\tau_\mathrm{eff}$ that has been taken is:

\begin{equation}
\tau_\mathrm{eff} = 0.0036 \left(\frac{\lambda_\mathrm{obs}}{1216
\mathrm{\AA}} \right)^{3.46}
\label{tau}
\end{equation}

\noindent
\citep{pre93,mad95}. Because the \lya\ line of an object at $z = 3.13$
falls in the blue wing of the $V$ filter, the fraction of the continuum flux
falling in the $V$ filter that is absorbed is small and $Q$ is near
unity: $Q_\mathrm{V} \sim 0.97$. In the narrow-band, $Q_\mathrm{nb} =
0.92$ for a source at $z = 3.13$.  

To calculate the equivalent width of an individual \lya\ line, Eqs.\
(\ref{fnb}) -- (\ref{fi}) were solved for $\beta$, $C$ and
$F_{\mathrm{Ly}\alpha}$. This was done in the following way. Equation
(\ref{fnb}) was multiplied by $\Delta \lambda_\mathrm{nb}^\prime =
\Delta \lambda_\mathrm{nb} / \epsilon_\mathrm{nb}$ and Eq.\ (\ref{fv})
by $\Delta \lambda_\mathrm{V}^\prime = \Delta \lambda_\mathrm{V} /
\epsilon_\mathrm{V}$. Substituting $C = f_\mathrm{\lambda,I} /
\lambda_\mathrm{eff,I}^{\beta}$ (Eq.\ \ref{fi}) gave:

\begin{eqnarray}
\label{fnb2}
\Delta \lambda_\mathrm{nb}^\prime\,f_{\lambda,\mathrm{nb}} =
f_{\lambda,\mathrm{I}}\,\Delta \lambda_\mathrm{nb}^\prime\,Q_\mathrm{nb}
\left(\frac{\lambda_\mathrm{eff,nb}}{\lambda_\mathrm{eff,I}}
\right)^\beta + F_{\mathrm{Ly}\alpha} \\
\label{fv2}
\Delta \lambda_\mathrm{V}^\prime\,f_{\lambda,\mathrm{V}} =
f_{\lambda,\mathrm{I}}\,\Delta \lambda_\mathrm{V}^\prime\,Q_\mathrm{V}
\left(\frac{\lambda_\mathrm{eff,V}}{\lambda_\mathrm{eff,I}}
\right)^\beta + F_{\mathrm{Ly}\alpha}. 
\end{eqnarray} 

Subtraction of Eq.\ (\ref{fnb2}) from Eq.\ (\ref{fv2})
results in an equation of the form $a^\beta - b^\beta =$ {\em
constant}. This equation was solved numerically.

When $\beta$ was computed, the UV continuum flux density $C$ and the \lya\
line flux $F_\mathrm{Ly\alpha}$ were calculated using Eqs.\
(\ref{fnb}) and (\ref{fv}): 

\begin{equation}
C = \frac{\Delta \lambda_\mathrm{V}^\prime f_{\lambda,\mathrm{V}} -
  \Delta \lambda_\mathrm{nb}^\prime f_{\lambda,\mathrm{nb}}}{\Delta
  \lambda_\mathrm{V}^\prime Q_\mathrm{V}
  \lambda_\mathrm{eff,V}^\beta - \Delta \lambda_\mathrm{nb}^\prime
  Q_\mathrm{nb} \lambda_\mathrm{eff,nb}^\beta}
\end{equation}

\noindent
and

\begin{equation}
F_\mathrm{Ly\alpha} = \frac{f_{\lambda,\mathrm{nb}} / (Q_\mathrm{nb}
  \,\lambda_\mathrm{eff,nb}^\beta) - f_{\lambda,\mathrm{V}} /
  (Q_\mathrm{V} \,\lambda_\mathrm{eff,V}^\beta)}{1/(\Delta
  \lambda_\mathrm{nb}^\prime \, Q_\mathrm{nb} \,
  \lambda_\mathrm{eff,nb}^\beta) - 1/(\Delta \lambda_\mathrm{V}^\prime
  \,Q_\mathrm{V} \, \lambda_\mathrm{eff,V}^\beta)}.
\end{equation}

\noindent
With $C$ and $F_\mathrm{Ly\alpha}$, the equivalent width ({\em EW}) for each
object was computed:

\begin{equation}
\label{ewus}
EW_{\mathrm{obs}} = \frac{F_\mathrm{Ly\alpha}}{C
  (\lambda_\mathrm{Ly\alpha} (1+z))^\beta}
\end{equation}

\noindent
with $\lambda_\mathrm{Ly\alpha}$ the wavelength of the \lya\ line. The
rest frame equivalent width ({\em EW}$_0$) is given by: {\em EW}$_0 =$
{\em EW}$_{\mathrm{obs}} / (1+z)$.

\begin{figure}[!t]
\resizebox{\hsize}{!}{\includegraphics{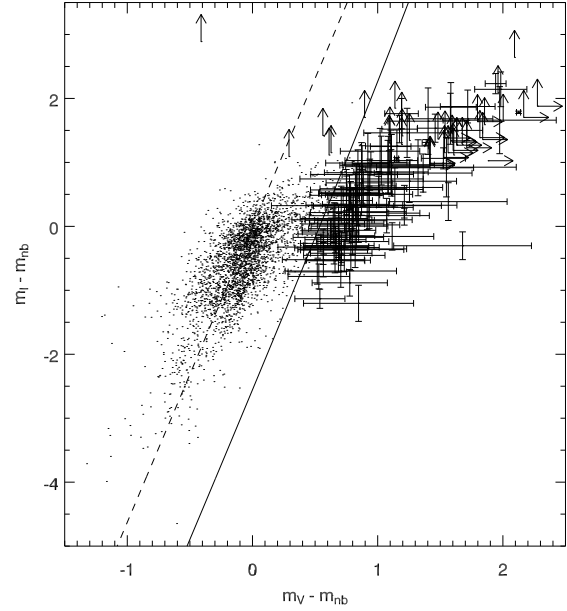}}
\caption{Color--color diagram for the 3209 objects detected in the
narrow-band image with a signal-to-noise greater than 5. The dashed
line shows the color of objects with a rest-frame equivalent width of
{\em EW}$_0 = 0$ \AA. The solid line indicates where {\em EW}$_0 = 15$ \AA.
Objects not detected in the $V$-band and/or in the $I$-band are
plotted with an arrow.}
\label{colorplot}
\end{figure}

To estimate the uncertainties in the computed parameters, the observed
flux densities were randomly varied 50\,000 times over a range having
a standard deviation equal to the uncertainty. The distributions of
$\beta$, $C$, $F_\mathrm{Ly\alpha}$ and {\em EW}$_0$ were used to estimate
the errors in these quantities. Because the values of the equivalent
width were not normally-distributed (Gaussian) around the central
value, two errors were calculated, labelled $\Delta EW_0^-$
and $\Delta EW_0^+$. These were computed from the
values in the distribution that were outside the central 99.73\% of
all values. The difference between these values and the central value
was taken as defining three sigma uncertainties.

For each object detected in the narrow-band image, the line flux, UV
continuum and equivalent width and their errors were computed.
Because no $I$-band data had been taken yet at the time that the
candidates for the spectroscopy had to be selected, a flat spectrum
($\beta = -2$) was assumed for all sources. In Fig.\ \ref{colorplot},
the $m_\mathrm{I}-m_\mathrm{nb}$ color is plotted against the
$m_\mathrm{V}-m_\mathrm{nb}$ color. Following \citet{ven02}, objects
with {\em EW}$_0 > 15$ \AA\ and {\em EW}$_0/\Delta EW_0^- > 3$ were
selected as candidate \lya\ emitters. Each individual \lya\ candidate
was inspected visually in order to remove spurious candidates, like
leftover cosmic rays or objects in the ``spikes'' of bright stars. This
resulted in a list of 77 candidate \lya\ emitters with {\em EW}$_0 > 15$
\AA\ of which 6 had $15$ \AA\ $<$ {\em EW}$_0 < 20$ \AA. 

\begin{figure}[!t]
\resizebox{\hsize}{!}{\includegraphics{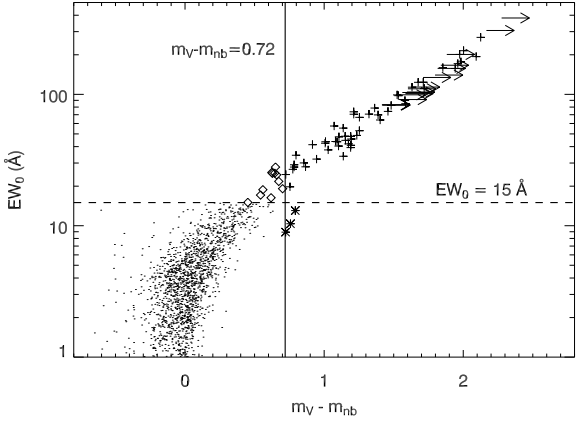}}
\caption{Equivalent width (calculated from Eqs.\
  \ref{fnb}--\ref{ewus}) versus $m_{\mathrm{V}}-m_{\mathrm{nb}}$
  color. For clarification, only objects with an {\em EW}$_0 > 1$ \AA\ are
  plotted and the error bars are left out. The solid line indicates a
  color of $m_{\mathrm{V}}-m_{\mathrm{nb}} = 0.72$, while the dashed
  line denotes the division between low ({\em EW}$_0 < 15$ \AA) and high
  ({\em EW}$_0 > 15$ \AA) equivalent width objects.}
\label{nbvewplot}
\end{figure}

The main difference between the usage of Eq.\ (\ref{ewbun}) and Eqs.\
(\ref{fnb})--(\ref{ewus}) to compute the equivalent width can be seen
in Fig.\ \ref{nbvewplot}. Using Eq.\ (\ref{ewbun}) (and thereby
assuming a fixed slope of $\beta = -2$) the line with {\em EW}$_0$ = 15 \AA\
would lie at a constant $m_V - m_{\mathrm{nb}} = 0.72$ (the solid line
in Fig.\ \ref{nbvewplot}). As a result, the {\em EW}$_0$ of three very blue
objects (with $\beta < -2$) would be overpredicted, falsely selecting
objects as \lya\ emitters (e.g.\ the three crosses in Fig.\
\ref{nbvewplot}). On the other hand, 11 red \lya\ emitters with
$\beta > -2$ would not pass the selection criterion $m_V -
m_{\mathrm{nb}} > 0.72$, while their {\em EW}$_0$ as calculated with Eqs.\
(\ref{fnb})--(\ref{ewus}) is greater than 15 \AA\ (diamonds in Fig.\
\ref{nbvewplot}).

\section{Spectroscopy}
\label{spec}

\subsection{Spectroscopic observations}
\label{specobs}

Spectra of candidate \lya\ emitters were taken during three separate
observing sessions (see Table 1 for an overview). The first
spectroscopy session was carried out on 2001, September 22 with
VLT/FORS2 in the multi-object spectroscopy mode with 19 movable slits
of a fixed length of $\sim 20-22$\arcsec. The night was photometric,
but because of strong winds ($>$ 12 m s$^{-1}$) the seeing fluctuated
between 1\arcsec\ and 2\arcsec. Spectra of 12 candidate \lya\ emitters
were obtained for $4 \times 2700$ s and $1 \times 1800$ s through
1\arcsec\ slits with the 1400V grism at a dispersion of 0.5
\AA\,pixel$^{-1}$. This grism was chosen for a number of
reasons. First, it has a high peak efficiency of $\sim$85\% at
wavelengths that corresponds to the redshifted \lya\ line of the radio
galaxy. Secondly, because observations of high redshift \lya\ emitting
galaxies have shown that the width of the \lya\ line lies
predominantly in the range 200--500 \kms\
\citep[e.g.][]{pen00a,daw02,hu04}, the resolution of the grism ($R = 2100$,
corresponding to $\sim 150$ \kms) ensured that the \lya\ emission line
is marginally resolved (see Sect.\ \ref{spectra}), maximizing the
signal-to-noise of the observed line. Also, the resolution is large
enough to distinguish a high redshift \lya\ emitting galaxy from a low
redshift contaminant, the [\ion{O}{ii}] $\lambda\lambda 3726,3729$
emitter. With the 1400V grism the [\ion{O}{ii}] doublet is resolved
(see Fig.\ \ref{oii} for two examples). For the wavelength calibration
exposures of He, HgCd and Ne arc lamps were obtained. The
spectrophotometric standard star LTT 1788 \citep{sto83,bal84} was
observed with a 5\arcsec\ slit for the flux calibration.

On 2001, October 18, 19 and 20 spectra were obtained with FORS2 in the
mask multi-object spectroscopy mode. In this mode objects are observed
through a user defined, laser-cut mask with slits which had variable
lengths (typically $10-12$\arcsec) and widths of 1\arcsec. The nights
were photometric with an average seeing of 1\arcsec. The 1400V grism
was used to observe 37 candidate \lya\ emitters in two masks, of which
25 were included in both masks. The first mask was observed for $4
\times 2700$ s and the second mask for $10 \times 2700$ s and $1
\times 2100$ s. The pixels were binned by $2 \times 2$ to avoid the
noise in the spectra being dominated by read noise. This resulted in a
dispersion of 1 \AA\,pixel$^{-1}$ and a spatial scale of 0\farcs4
pixel$^{-1}$. Spectra of the standard star LTT 1788 were obtained for
the flux calibration.

During the last observing session (2001, November 15 and 16), the
instrument used was FORS1 on Melipal (VLT UT3). The main goal of this
run was to measure the polarization of the radio galaxy (C. De Breuck
et al., in preparation). Due to constraints on the positioning and
orientation of the mask, only three candidate \lya\ emitters could be
observed. The width of the slits was 1\arcsec. The total exposure time
was 19\,800 s. The average seeing of these photometric nights was
0\farcs8. The grism used for the observations was the ``300V'' with a
resolution of 440, a dispersion of 2.64 \AA\,pixel$^{-1}$ and a
spatial scale of 0\farcs2 pixel$^{-1}$. The spectrophotometric
standard stars Feige 110 and LTT 377 \citep{sto83,bal84} were observed
for the flux calibration.

\subsection{Data reduction}

The spectra were reduced in the following way. Individual frames were
flat-fielded using lamp flats, cosmic rays were identified and removed
and the background was subtracted. The next step was the extraction of
the one-dimensional (1D) spectra. Typical aperture sizes were
1\arcsec--1\farcs5. If the spectrum of the object could
be seen in the individual frames, then a spectrum was extracted from
each frame and these spectra were combined. If the object was
undetected in the individual frames, then the background subtracted
two-dimensional frames were combined and a 1D spectrum was extracted
from this image. All 1D spectra were wavelength calibrated using the
arc lamp spectra. For spectra taken with the 1400V grism the rms of
the wavelength calibration was always better than 0.05 \AA, which
translates to $\Delta z = 0.00004$ at $z \sim 3.13$. The wavelength
calibration with the 300V grism had an rms of 0.8 \AA\ ($\Delta z =
0.0007$ at $z \sim 3.13$). A heliocentric correction was applied on
measured redshifts to correct for the radial velocity of the Earth in
the direction of the observations. Finally, the spectra were flux
calibrated. The fluxes of the photometric standard stars in the
individual images were consistent with each other to within 5\%, so we
estimate that the flux calibration of the spectra is accurate to $\sim
5$\%. 

\subsection{Results}

\begin{table*}
\begin{center}
\caption{\label{ztable} Position and properties of the \lya\ emission
  line of the 31 confirmed \lya\ emitters and the radio galaxy.}
\begin{tabular}{cccccccc}
\hline
\hline
Object & \multicolumn{2}{c}{Position} & $z$ & Flux & {\em EW}$_0$ &
{\em FWHM} & {\em SFR}$_{\rm Ly\alpha}$ \\ 
{} & $\alpha_{\mathrm{J}2000}$ & $\delta_{\mathrm{J}2000}$ & {} &
$10^{-17}$ \ergscm\ & \AA & \kms\ & \msunyr\ \\
\hline
344  & 03 18 05.83 & $-$25 37 55.7 & 3.1332 $\pm$ 0.0001 & 4.2
$\pm$ 0.3 & 144$^{+339}_{-30}$ & 260 $\pm$ 20 & 2.8 $\pm$ 0.2 \\
695  & 03 18 13.26 & $-$25 37 18.9 & 3.1712 $\pm$ 0.0001 & 2.3
$\pm$ 0.2 & 246$^{+553}_{-51}$ & 180 $ \pm$ 20 & 4.0 $\pm$ 0.4 \\
748  & 03 18 21.51 & $-$25 37 13.9 & 3.1359 $\pm$ 0.0003 & 0.4
$\pm$ 0.1 & 70$^{+949}_{-20}$ & 260 $\pm$ 50 & 1.1 $\pm$ 0.2 \\
995  & 03 18 09.67 & $-$25 36 47.5 & 3.1239 $\pm$ 0.0003 & 4.3
$\pm$ 0.6 & 65$^{+36}_{-14}$ & 450 $\pm$ 30 & 2.6 $\pm$ 0.4 \\
1029 & 03 18 00.07 & $-$25 36 43.6 & 3.1301 $\pm$ 0.0007 & 0.3
$\pm$ 0.1 & 39$^{+17}_{-8}$ & $<$ 590 & 1.7 $\pm$ 0.2 \\
1099 & 03 18 02.79 & $-$25 36 35.8 & 3.1313 $\pm$ 0.0004 & 1.1
$\pm$ 0.3 & 42$^{+65}_{-11}$ & 240 $\pm$ 70 & 1.4 $\pm$ 0.3 \\
1147 & 03 18 03.99 & $-$25 36 31.9 & 3.1666 $\pm$ 0.0003 & 6.0
$\pm$ 2.2 & 318$^{+1000}_{-90}$ & 290 $\pm$ 40 & 1.4 $\pm$ 0.2 \\
1203 & 03 18 12.55 & $-$25 36 23.0 & 3.1234 $\pm$ 0.0002 & 10.3
$\pm$ 1.8 & 78$^{+24}_{-12}$ & 220 $\pm$ 10 & 3.3 $\pm$ 0.3 \\
1361 & 03 18 01.57 & $-$25 36 07.8 & 3.1306 $\pm$ 0.0002 & 1.5
$\pm$ 0.2 & 106$^{+234}_{-24}$ & 210 $\pm$ 40 & 2.6 $\pm$ 0.3 \\
1395 & 03 18 24.78 & $-$25 36 05.7 & 3.1442 $\pm$ 0.0011 & 1.8
$\pm$ 0.6 & $\lesssim$ 30 & 790 $\pm$ 190 & 0.4 $\pm$ 0.1 \\
1446 & 03 18 04.89 & $-$25 35 57.7 & 3.1316 $\pm$ 0.0002 & 0.5
$\pm$ 0.1 & 12$^{+8}_{-4}$ & 210 $\pm$ 50 & 0.6 $\pm$ 0.2 \\
1498 & 03 18 14.14 & $-$25 35 54.3 & 3.1319 $\pm$ 0.0003 & 1.0
$\pm$ 0.1 & 58$^{+435}_{-18}$ & 460 $\pm$ 60 & 0.6 $\pm$ 0.1 \\
1518 & 03 18 16.78 & $-$25 35 46.2 & 3.1311 $\pm$ 0.0007 & 3.2
$\pm$ 0.8 & 23$^{+2}_{-2}$ & 570 $\pm$ 70 & 4.8 $\pm$ 0.2 \\
1612 & 03 18 03.49 & $-$25 35 39.2 & 3.1222 $\pm$ 0.0003 & 3.1
$\pm$ 0.3 & $>$ 370 & 500 $\pm$ 40 & 1.8 $\pm$ 0.3 \\
1710 & 03 18 15.17 & $-$25 35 28.9 & 3.1462 $\pm$ 0.0004 & 1.9
$\pm$ 0.3 & 61$^{+61}_{-14}$ & 420 $\pm$ 50 & 1.0 $\pm$ 0.1 \\
1724 & 03 18 06.92 & $-$25 35 26.2 & 3.1426 $\pm$ 0.0002 & 0.4
$\pm$ 0.1 & $>$ 150 & 160 $\pm$ 50 & 0.9 $\pm$ 0.1 \\
1753 & 03 18 01.22 & $-$25 35 22.3 & 3.1301 $\pm$ 0.0001 & 1.5
$\pm$ 0.2 & 32$^{+16}_{-7}$ & 130 $\pm$ 40 & 2.1 $\pm$ 0.4 \\
1759 & 03 18 25 50 & $-$25 35 20.4 & 3.1271 $\pm$ 0.0002 & 0.6
$\pm$ 0.1 & 48$^{+29}_{-11}$ & 120 $\pm$ 50 & 2.5 $\pm$ 0.4 \\
1829 & 03 18 15.12 & $-$25 35 13.1 & 3.1313 $\pm$ 0.0002 & 0.7
$\pm$ 0.1 & 24$^{+19}_{-7}$ & 200 $\pm$ 40 & 1.3 $\pm$ 0.3 \\
1867 & 03 18 09.00 & $-$25 34 59.6 & 3.1358 $\pm$ 0.0002 & 5.8
$\pm$ 0.4 & 56$^{+3}_{-3}$ & 650 $\pm$ 40 & 11.6 $\pm$ 0.2 \\
1891 & 03 18 04.29 & $-$25 35 04.9 & 3.1454 $\pm$ 0.0002 & 0.6
$\pm$ 0.1 & $\lesssim$ 31 & 130 $\pm$ 40 & 0.8 $\pm$ 0.2 \\
1946 & 03 18 20.29 & $-$25 34 59.4 & 3.1335 $\pm$ 0.0003 & 0.7
$\pm$ 0.1 & $\lesssim$ 19 & 260 $\pm$ 50 & 0.7 $\pm$ 0.3 \\
1955 & 03 18 07.58 & $-$25 34 55.5 & 3.1391 $\pm$ 0.0003 & 0.7
$\pm$ 0.1 & $>$ 139 & 330 $\pm$ 60 & 1.1 $\pm$ 0.2 \\
1962 & 03 18 12.03 & $-$25 34 52.8 & 3.1407 $\pm$ 0.0007 & 1.7
$\pm$ 0.4 & 33$^{+12}_{-6}$ & 600 $\pm$ 130 & 0.9 $\pm$ 0.1 \\
1968 & 03 18 05.14 & $-$25 34 51.2 & 3.1564 $\pm$ 0.0004 & 0.4
$\pm$ 0.1 & $\lesssim$ 66 & 300 $\pm$ 80 & 1.7 $\pm$ 0.3 \\
2413 & 03 18 20.40 & $-$25 33 00.3 & 3.1339 $\pm$ 0.0002 & 0.7
$\pm$ 0.2 & 62$^{+28}_{-11}$ & 200 $\pm$ 50 & 2.7 $\pm$ 0.3 \\
2487 & 03 17 59.63 & $-$25 34 01.3 & 3.1175 $\pm$ 0.0024 & 19.3
$\pm$ 2.6 & 202$^{+9}_{-8}$ & 2510 $\pm$ 190 & --$^a$ \\
2637 & 03 18 22.52 & $-$25 34 10.1 & 3.1224 $\pm$ 0.0002 & 1.3
$\pm$ 0.3 & 79$^{+43}_{-16}$ & 200 $\pm$ 50 & 3.2 $\pm$ 0.5 \\
2871 & 03 18 25.71 & $-$25 33 40.5 & 3.1334 $\pm$ 0.0003 & 0.4
$\pm$ 0.1 & $>$ 75 & 250 $\pm$ 70 & 0.7 $\pm$ 0.1 \\
3101 & 03 18 24.09 & $-$25 32 12.2 & 3.1313 $\pm$ 0.0003 & 4.3
$\pm$ 1.2 & 114$^{+47}_{-18}$ & 800 $\pm$ 100 & 5.2 $\pm$ 0.3 \\
3388 & 03 18 26.09 & $-$25 32 53.1 & 3.1379 $\pm$ 0.0002 & 2.1
$\pm$ 0.1 & 112$^{+26}_{-13}$ & 370 $\pm$ 20 & 8.6 $\pm$ 0.3 \\
\hline
HzRG & 03 18 12.01 & $-$25 35 10.8 & 3.1307 $\pm$ 0.0001 & 155.1
$\pm$ 4.2 & 257$^{+9}_{-8}$ & 1320 $\pm$ 10 & 76.1 $\pm$ 0.4$^a$ \\
\hline
\end{tabular}
\end{center}
$^a$ Object contains an AGN, therefore the estimated {\em SFR} is unreliable. 
\end{table*}

\begin{figure}[!t]
\resizebox{\hsize}{!}{\includegraphics{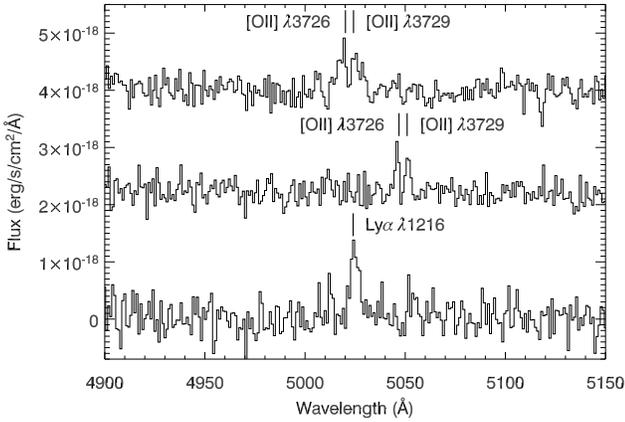}}
\caption{Spectra of two [\ion{O}{ii}] emitters observed in the field
  (top two spectra). For comparison, the spectrum of one of the
  confirmed \lya\ emitters is shown at the bottom. The spectra are
  offset from each other by $1.5 \times 10^{-18}$ erg s$^{-1}$
  cm$^{-2}$ \AA$^{-1}$.}
\label{oii}
\end{figure}

Spectra were obtained for a total of 40 candidate \lya\ emitters, of
which 11 were observed during two separate observing sessions, and
the central radio galaxy. Of the 40 candidate emitters, only 7 failed
to show an emission line. Six of these unconfirmed emitters had
predicted line fluxes below $10^{-17}$ \ergscm\ and were probably too
faint to be detected. Two of the 33 emission line objects showed two lines
with almost equal strength, separated by $\sim$4 \AA\ (Fig.\
\ref{oii}). These objects were identified to be [\ion{O}{ii}]
$\lambda\lambda 3726,3729$ emitters at a redshift of $\sim0.35$. One
of the [\ion{O}{ii}] emitters was re-observed in November 2001 and a
nearly flat continuum was revealed with no break around the emission
line, confirming that the object could not be a \lya\ emitter at $z
\sim 3.13$. None of the other emitters had more than one emission line
in the spectrum. This excluded identification of the emission line
with [\ion{O}{iii}] $\lambda 5007$, because then the confirming
[\ion{O}{iii}] $\lambda 4959$ would have been visible. Furthermore, a
number of emitters showed an asymmetric line profile (Figs.\
\ref{spec344995}--\ref{spec33881687}), a feature often seen in spectra
of high redshift \lya\ emitters
\citep[e.g.][]{aji02,daw02}. Therefore, the 31 remaining emitters were
identified with being \lya\ emitters. The fraction of contaminants in
our sample is 2/33 = 6.1\%, similar to the fraction of low redshift
interlopers of 6.5\% in the study of \lya\ emitters at $z\sim3.09$ of
\citet{ste00}.

\section{Properties of the \lya\ emitting galaxies}
\label{properties}

The one-dimensional \lya\ emission lines were fitted by a Gaussian
function and -- if absorption was clearly present -- in combination
with Voigt absorption profiles. The best fit Gaussian was used to
calculate the redshift, line flux and {\em FWHM} of each emitter. In Table
2 the properties of the confirmed \lya\ emitters are
summarized. The IDs correspond to the object's number in the
SExtractor catalog. The rest-frame equivalent width {\em EW}$_0$ was taken
from the imaging. The star formation rate ({\em SFR}) was calculated using
the \lya\ line flux derived from the images, and assuming Case B
recombination and using the H$\alpha$ luminosity to {\em SFR} conversion
from \citet[][see Sect.\ \ref{sfr}]{mad98}.

\subsection{Line profiles}
\label{spectra}

\begin{table}
\begin{center}
\caption{Characteristics of the Voigt absorption
  profiles derived from the spectra. For each absorption profile, its
  centre relative to the peak of the emission line, width ($b$) and
  \ion{H}{i} column density ($N$) is printed.}
\begin{tabular}{cccc}
\hline
\hline
Object & Centre & $b$ (\kms) & log $N$ (cm$^{-2}$) \\
\hline
344 & $-80 \pm 10$ & 52 $\pm$ 11 & 14.4 $\pm$ 0.1 \\ 
995 & $-150 \pm 20$ & 74 $\pm$ 59 & 16.0 $\pm$ 2.5 \\
1147 & $-70 \pm  20$ & 101 $\pm$ 24 & 14.6 $\pm$ 0.1 \\ 
1203 & $-90 \pm 10$ & 104 $\pm$ 11 & 14.9 $\pm$ 0.1 \\ 
1518 & $80 \pm 50$ & 38 $\pm$ 16 & 13.8 $\pm$ 0.2 \\
 {}  & $-210 \pm 60$ & 155 $\pm$ 52 & 15.0 $\pm$ 0.2 \\
1612 & $-150 \pm 10$ & 79 $\pm$ 16 & 14.4 $\pm$ 0.1 \\
1710 & $-130 \pm 20$ & 80 $\pm$ 30 & 14.1 $\pm$ 0.2  \\ 
1867 & $-60 \pm 20$ & 30 $\pm$ 16 & 13.1 $\pm$ 0.3 \\
2487 & $250 \pm 170$ & 108 $\pm$ 14 & 14.6 $\pm$ 0.1 \\
 {}  & $-1150 \pm 200$ & 629 $\pm$ 93 & 16.0 $\pm$ 0.2 \\
3101 & $-40 \pm 110$ & 193 $\pm$ 131 & 14.4 $\pm$ 0.4 \\
 {}  & $-240 \pm 20$ & 79 $\pm$ 38 & 14.3 $\pm$ 0.3 \\ 
3388 & $-130 \pm 10$ & 59 $\pm$ 13 & 14.3 $\pm$ 0.1 \\
HzRG & $200 \pm 10$ & 151 $\pm$ 9 & 14.9 $\pm$ 0.1 \\
{} & $-270 \pm 10$ & 245 $\pm$ 25 & 14.9 $\pm$ 0.1 \\
{} & $-660 \pm 10$ & 144 $\pm$ 20 & 14.8 $\pm$ 0.1 \\
{} & $-970 \pm 20$ & 131 $\pm$ 27 & 14.2 $\pm$ 0.1 \\
\hline
\end{tabular}
\end{center}
\end{table}

As mentioned in the previous paragraph, emitters which clearly showed
an emission line with an absorbed blue wing (see Figs.\
\ref{spec344995}--\ref{spec33881687}) were fitted by a Gaussian
emission line with one or more Voigt absorption profiles. The
characteristics of the absorption profiles are listed in Table
3. The other emitters were fitted by a single Gaussian. The
observed width of the line was deconvolved with the instrumental
width, which was 150 \kms\ (see Sect.\ \ref{specobs}).

\begin{figure}
\resizebox{\hsize}{!}{\includegraphics{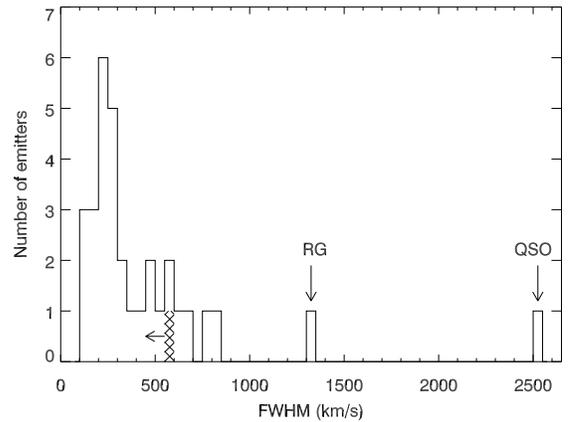}}
\caption{Histogram of the line widths of the emitters.}
\label{linefwhm}
\end{figure}

The radio galaxy has an emission line that can be fitted by a Gaussian
with a {\em FWHM} of $\sim1300$ \kms. The line width is very similar to that
of other HzRGs \citep[e.g.][]{deb01,wil02}. Only one of the confirmed
emitters has a broad \lya\ line. Emitter \#2487 has a line {\em FWHM} of
$\sim$2500 \kms, and is therefore likely to also harbour an AGN. The
{\em FWHM} of the Ly$\alpha$ emission line of the rest of the emitters
ranges from 120 \kms\ to 800 \kms\ (Fig.\ \ref{linefwhm}), with a
median of 260 \kms\ and a mean of 340 \kms.

The inferred column densities of the absorbers are in the range of
$10^{13}$--$10^{16}$ cm$^{-2}$ (see Sect.\ \ref{notes}). Using the
spatial extend in the 2D spectra as an estimate of the size of the
\ion{H}{i} absorber, the amount of projected neutral HI near the
emitters is in the range of $> 2 \times 10^2 - 5 \times 10^4$
{\em M}$_{\sun}$ (see Sect.\ \ref{notes} for the details of the individual
emitters). For the fainter \lya\ emitters, it cannot be excluded that
the troughs are due to substructure in the \lya\ emitting regions,
rather than \ion{H}{i} absorption.

\subsection{Continuum colors}
\label{colors}

\begin{figure}[!t]
\resizebox{\hsize}{!}{\includegraphics{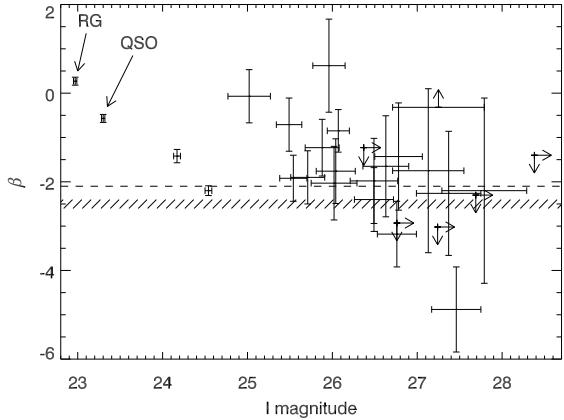}}
\caption{$I$ magnitude versus UV continuum slope $\beta$. The dashed
  line corresponds to $\beta = -2.1$, the color of an unobscured
  continuously star forming galaxy with an age of $\sim 10^8 - 10^9$
  yr. The hashed area indicates the color of a young
  ($\sim 10^6$ yr), star forming galaxy.}
\label{ivsbeta}
\end{figure}

In Table 4, the magnitudes and UV continuum slopes of
the confirmed emitters are listed (for a description on how the slope
was calculated, see Sect.\ \ref{selection}). For objects detected in
the ACS image which were not resolved into several components, the
$I_{814}$ magnitude and the continuum slope $\beta_{\mathrm{ACS}}$
calculated using this magnitude is given. In Fig.\ \ref{ivsbeta} the
$I$ magnitude is plotted against the continuum slope. Excluding the
radio galaxy and emitter \#2487, which contains an AGN (Sect.\
\ref{spectra}), the UV continuum slope of the confirmed emitters
ranges from $\beta = 0.62$ to $\beta = -4.88$ with a median of $\beta
= -1.76$.

The blue median color of the \lya\ emitters may be due to a selection
effect. The candidate emitters for spectroscopy were selected when
only one broad-band flux was available and a slope of $\beta = -2$ was
assumed for all objects to compute the equivalent width (see Sect.\
\ref{selection} and Fig.\ \ref{nbvewplot}). Because the narrow-band is
on the blue side of the broad-band filter that was used, the
equivalent width and line flux of bluer objects with $\beta < -2$ tend
to have been overestimated, while ``red'' objects ($\beta > -2$) have
an equivalent width and \lya\ flux that are likely to be
underestimated (see Sect.\ \ref{selection}). This effect could have
biased the spectroscopic sample towards blue objects. For example, the
emitter with the bluest color, \#1446, has an equivalent width of {\em
EW}$_0 = 12$ \AA, which falls below the selection criteria ({\em
EW}$_0 > 15$ \AA), but the object was selected for spectroscopy
because it had $m_\mathrm{V}-m_\mathrm{nb} > 0.72$ (see discussion at
the end of Sect.\ \ref{selection}). To determine the effect of this
bias, the color of a flux limited sample was determined. There are 31
(candidate) \lya\ emitting objects with a \lya\ flux $> 1.5 \times
10^{-17}$ \ergscm\ in the field, of which 25 are confirmed. Again
excluding the radio galaxy and the AGN, the median color of the
remaining 29 emitters is $\beta = -1.70$. This is bluer than the
average color of \lya\ emitting Lyman Break Galaxies (LBGs), which
have a slope of $\beta = -1.09 \pm 0.05$ \citep{sha03}.

\begin{table*}
\begin{center}
\caption{\label{magnitudes} Narrow-band, $V$, $I$ and ACS $I_{814}$
  magnitudes and UV continuum slopes $\beta$ of the confirmed
  emitters. Colors were measured in a circular aperture, while
  elliptical apertures were used to determine the total magnitudes
  (see Sect.\ \ref{photometry}). For objects with a signal-to-noise less
  than two a 2 $\sigma$ upper limit is given.}
\begin{tabular}{ccccccc}
\hline
\hline
Object & NB magnitude & $V$ magnitude & $I$ magnitude & $I_{814}$
magnitude & $\beta$ & $\beta_\mathrm{ACS}$  \\
\hline
344 & 24.63 $\pm$ 0.06 & 26.62 $\pm$ 0.22 & 26.78 $\pm$ 0.28 & 27.76
$\pm$ 0.23 & $-$1.43 $\pm$ 1.21 & $-$3.91 $\pm$ 1.06 \\
695 & 24.89 $\pm$ 0.09 & 26.42 $\pm$ 0.20 & 25.96 $\pm$ 0.19 & 26.53
$\pm$ 0.10 & 0.62 $\pm$ 1.05 & $-$0.92 $\pm$ 1.09 \\
748 & 25.45 $\pm$ 0.14 & 27.03 $\pm$ 0.37 & 27.13 $\pm$ 0.42 & $-$ & 
$-$1.75 $\pm$ 1.85 & $-$ \\ 
995 & 24.97 $\pm$ 0.08 & 26.14 $\pm$ 0.16 & 26.49 $\pm$ 0.23 & 26.59
$\pm$ 0.11 & $-$2.40 $\pm$ 0.72 & $-$2.65 $\pm$ 0.63 \\
1029 & 25.00 $\pm$ 0.07 & 26.19 $\pm$ 0.12 & 26.76 $\pm$ 0.23 & $-$ &
$-$3.18 $\pm$ 0.74 & $-$ \\
1099 & 25.23 $\pm$ 0.13 & 26.39 $\pm$ 0.28 & 26.49 $\pm$ 0.28 & $-$ &
$-$1.98 $\pm$ 0.96 & $-$ \\
1147 & 25.74 $\pm$ 0.12 & 27.83 $\pm$ 0.44 & $>$ 28.38 & $-^b$ &
$<$ $-$1.40 & $-$ \\ 
1203 & 24.79 $\pm$ 0.06 & 26.00 $\pm$ 0.11 & $-^a$ & 27.30 $\pm$ 0.15 &
$-$ & $-$4.77 $\pm$ 0.55 \\ 
1361 & 24.76 $\pm$ 0.10 & 26.47 $\pm$ 0.24 & 26.63 $\pm$ 0.27 & $-$ & 
$-$1.65 $\pm$ 1.14 & $-$ \\
1395 & 26.40 $\pm$ 0.21 & 27.50 $\pm$ 0.47 & $>$ 27.69 & $-$ & $<$
$-$2.30 & $-$ \\ 
1446 & 25.53 $\pm$ 0.10 & 26.32 $\pm$ 0.12 & 27.46 $\pm$ 0.29 & 26.55 $\pm$
0.12 & $-$4.88 $\pm$ 0.96 & $-$2.46 $\pm$ 0.46 \\ 
1498 & 26.14 $\pm$ 0.14 & 27.55 $\pm$ 0.29 & 27.79 $\pm$ 0.50 & 27.12
$\pm$ 0.18 & $-$2.20 $\pm$ 2.09 & $-$0.49 $\pm$ 1.06 \\
1518 & 23.65 $\pm$ 0.02 & 24.47 $\pm$ 0.03 & 24.54 $\pm$ 0.04 & 24.57
$\pm$ 0.03 & $-$2.20 $\pm$ 0.11 & $-$2.12 $\pm$ 0.12 \\
1612 & 25.79 $\pm$ 0.16 & 27.65 $\pm$ 0.54 & $>$ 27.38 & 27.84 $\pm$
0.28 & $-$ & $>$ $-$0.65 \\
1710 & 25.51 $\pm$ 0.08 & 27.09 $\pm$ 0.19 & 27.37 $\pm$ 0.38 & 27.34
$\pm$ 0.19 & $-$2.26 $\pm$ 1.40 & $-$2.18 $\pm$ 0.78 \\  
1724 & 25.74 $\pm$ 0.15 & $>$ 27.91 & $>$ 27.44 & $-^b$ & $-$ & $-$ \\ 
1753 & 24.72 $\pm$ 0.09 & 25.67 $\pm$ 0.18 & 25.71 $\pm$ 0.20 & $-^c$
& $-$1.90 $\pm$ 0.60 & $-$\\ 
1759 & 24.77 $\pm$ 0.09 & 25.87 $\pm$ 0.19 & 26.02 $\pm$ 0.27 & $-$ &
$-$2.03 $\pm$ 0.83 & $-$ \\ 
1829 & 25.11 $\pm$ 0.13 & 25.73 $\pm$ 0.26 & 25.02 $\pm$ 0.25 & $-^c$
& $-$0.07 $\pm$ 0.60 & $-$\\
1867 & 22.86 $\pm$ 0.01 & 24.25 $\pm$ 0.03 & 24.17 $\pm$ 0.04 & $-^c$
& $-$1.42 $\pm$ 0.15 & $-$ \\ 
1891 & 25.56 $\pm$ 0.13 & 26.76 $\pm$ 0.29 & $>$ 27.24 & 27.45 $\pm$
0.20 & $<$ $-$3.02 & $-$3.50 $\pm$ 1.01 \\ 
1946 & 25.58 $\pm$ 0.18 & 26.40 $\pm$ 0.36 & $>$ 26.76 & $-$ &
$<$ $-$2.93 & $-$ \\ 
1955 & 25.59 $\pm$ 0.13 & 27.56 $\pm$ 0.47 & 27.25 $\pm$ 0.54 & 27.50 
$\pm$ 0.18 & $>$ $-$0.32 & $>$ $-$1.03 \\
1962 & 25.44 $\pm$ 0.08 & 26.44 $\pm$ 0.11 & 26.07 $\pm$ 0.13 & 26.10
$\pm$ 0.08 & $-$0.85 $\pm$ 0.48 & $-$0.99 $\pm$ 0.38 \\
1968 & 25.00 $\pm$ 0.14 & 26.48 $\pm$ 0.45 & $>$ 26.37 & 26.61 $\pm$
0.12 & $<$ $-$1.23 & $-$1.89 $\pm$ 1.63 \\ 
2413 & 24.51 $\pm$ 0.07 & 25.97 $\pm$ 0.16 & 26.04 $\pm$ 0.23 & $-$ &
$-$1.76 $\pm$ 0.73 & $-$ \\
2487 & 22.25 $\pm$ 0.01 & 23.40 $\pm$ 0.02 & 23.30 $\pm$ 0.02 & $-$ &
$-$0.57 $\pm$ 0.09 & $-$ \\
2637 & 24.88 $\pm$ 0.09 & 25.97 $\pm$ 0.17 & 25.88 $\pm$ 0.20 & $-$ &
$-$1.23 $\pm$ 0.64 & $-$ \\
2871 & 26.04 $\pm$ 0.18 & $>$ 27.64 & $>$ 27.22 & $-$ & $-$ & $-$ \\
3101 & 23.99 $\pm$ 0.04 & 25.70 $\pm$ 0.12 & 25.49 $\pm$ 0.15 & $-$ &
$-$0.71 $\pm$ 0.60 & $-$ \\
3388 & 23.31 $\pm$ 0.03 & 25.25 $\pm$ 0.09 & 25.54 $\pm$ 0.16 & $-$ &
$-$1.92 $\pm$ 0.52 & $-$ \\
\hline
HzRG & 21.19$^d$ $\pm$ 0.01 & 23.31$^d$ $\pm$ 0.02 & 22.97$^d$ $\pm$
0.02 & $-^c$ & 0.27 $\pm$ 0.09 & $-$ \\ 
\hline
\end{tabular}
\end{center}
$^a$ Photometry unreliable due to nearby bright star. \\
$^b$ Undetected in the ACS image, $I_{814} > 27.1$ mag\,arcsec$^{-2}$ \\
$^c$ Resolved by {\em HST} into several components. \\ 
$^d$ Photometry influenced by nearby objects.\\
\end{table*}

Models of galaxies with active star formation predict UV continuum
slopes in the range $\beta = -2.6$ to $\beta = -2.1$ for an
unobscured, continuously star forming galaxy with ages between a few
Myr and more than a Gyr \citep{lei99}. 18 out of the 27 (67\%)
confirmed \lya\ emitters for which $\beta$ could be measured, have
colors within 1 $\sigma$ consistent with being an unobscured starburst
galaxy. Of those, 15 (56\% of the sample) have such blue colors with 1
$\sigma$ that they could be star forming galaxies with ages of order
$10^6$ yr, which have $\beta \sim -2.5$.

\subsection{Morphologies}
\label{morphsec}

\begin{figure}[!t]
\includegraphics[width=8.5cm]{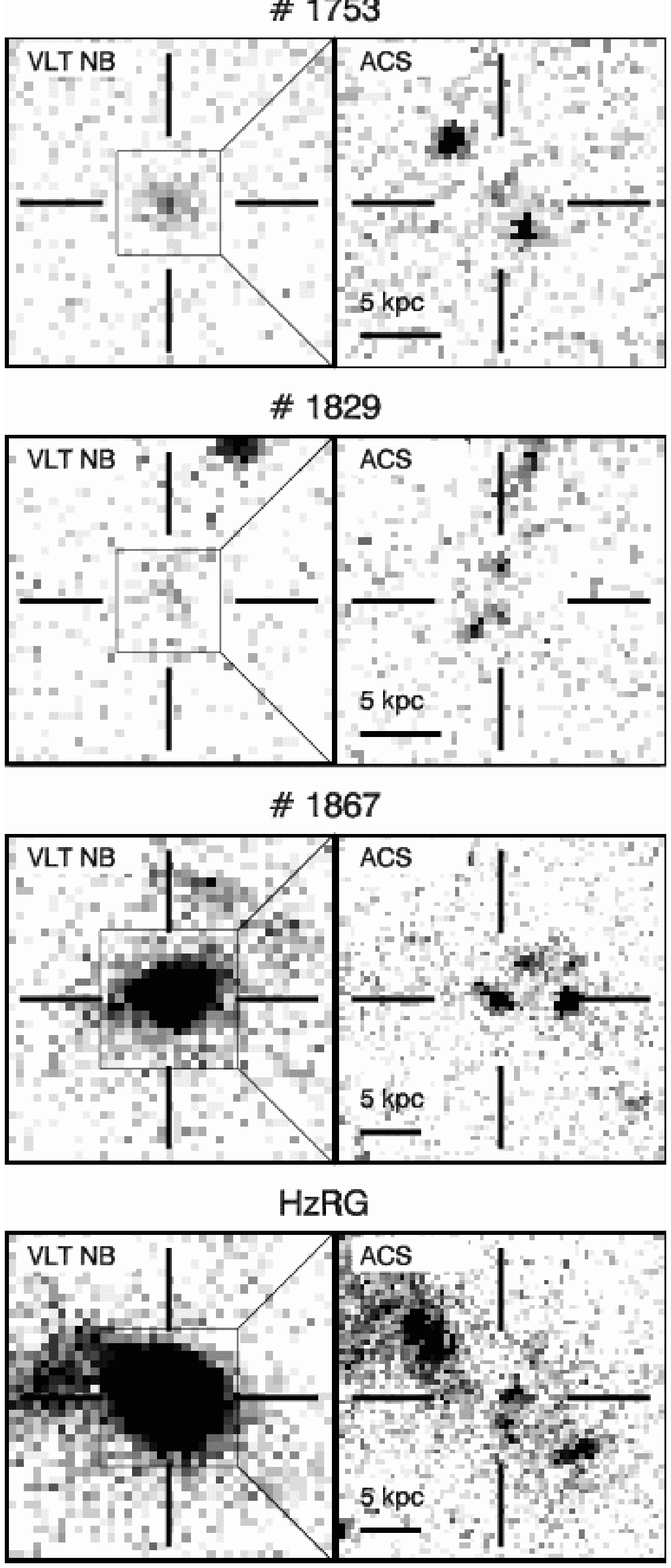} 
\caption{\label{multiple} VLT narrow-band and ACS images of \lya\
  emitters with a clumpy counterpart in the ACS image. The cutouts of
  the narrow-band image are $\sim 8$\arcsec\ on the side. The ACS
  images are zoomed in on the centre of the narrow-band image. The
  grayscale ranges from 0.5 to 5 times the rms background noise. The
  emission to the left of the radio galaxy in the VLT image (and in
  the top-left corner of the ACS image) are from a foreground galaxy
  at $z \sim 0.87$ (see also Sect.\ \ref{notes}).}
\end{figure}

\begin{figure}[!t]
\includegraphics[width=8.5cm]{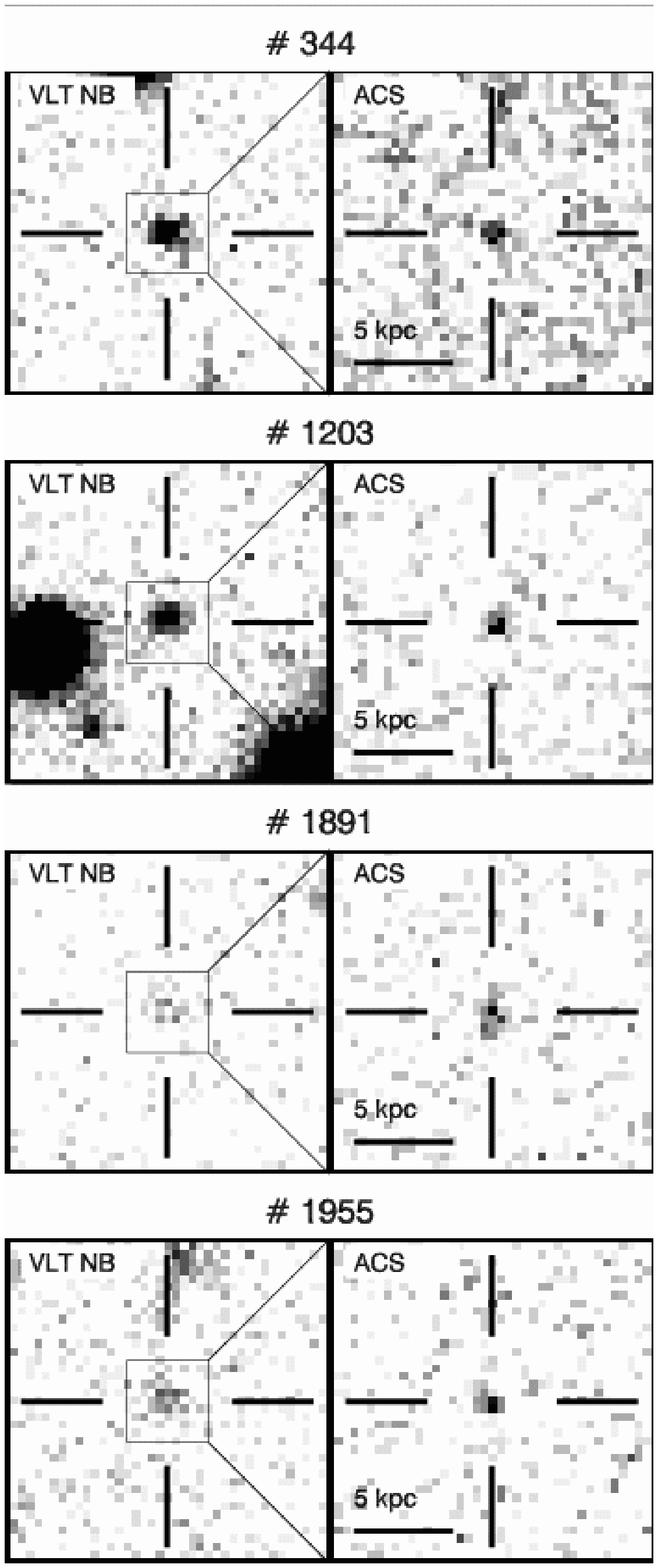} \\
\caption{\label{unresolved} VLT narrow-band and ACS images of \lya\
  emitters that remained unresolved in the ACS image. The cutouts of
  the narrow-band image are $\sim 8$\arcsec\ on the side. The
  grayscale ranges from 0.5 to 5 times the rms background noise.}
\end{figure}

\begin{figure*}[!ht]
\includegraphics[width=17cm]{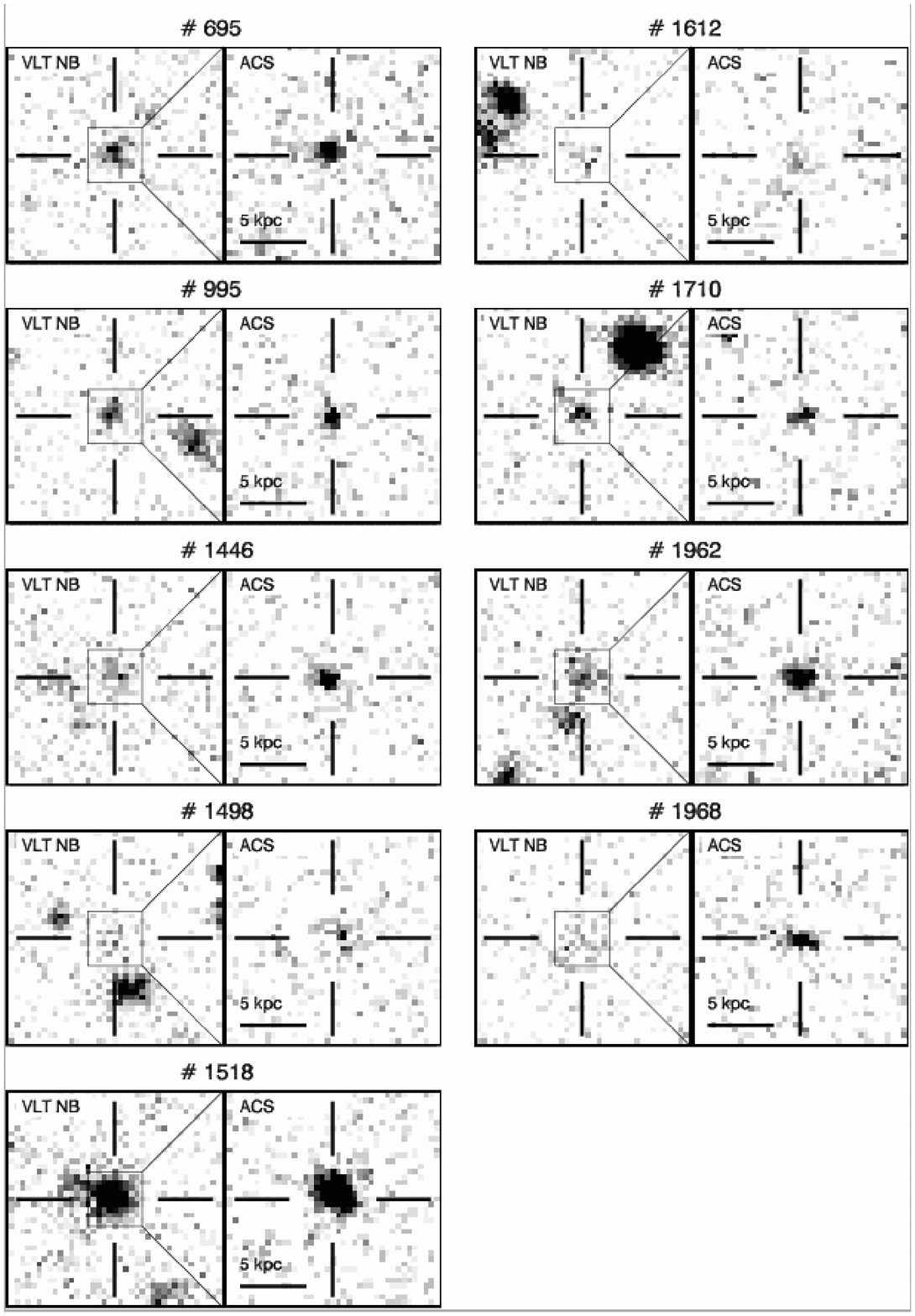} 
\caption{\label{resolved} VLT narrow-band and ACS images of \lya\
  emitters, which are resolved by the {\em HST}. The cutouts of the
  narrow-band image are $\sim 8$\arcsec\ on the side. The grayscale
  ranges from 0.5 to 5 times the rms background noise.}
\end{figure*}

Of the 32 confirmed \lya\ emitting sources, 19 (including the radio
galaxy) were located in the area that was imaged by the ACS (Fig.\
\ref{skydist}). Two of these emitters remained undetected to a depth
of $I_{814} >$ 27.1 mag\,arcsec$^{-2}$ (3 $\sigma$). On the position of the
radio galaxy, the ACS image shows several objects within 3\arcsec\
($\sim$25 kpc), surrounded by low surface brightness emission
($\gtrsim 24.8$ mag\,arcsec$^{-2}$, see Fig.\ \ref{multiple}). Such a
clumpy structure is often seen at the position of HzRGs
\citep[e.g.][]{pen99}. Interestingly, there are three other \lya\
emitters with morphologies that resemble the radio galaxy (Fig.\
\ref{multiple}). Each of these three objects consists of at least
three clumps of emission, which are less than one kpc separated from
each other. The remainder of the confirmed emitters can be identified
with single objects in the ACS image.

\begin{figure}
\resizebox{\hsize}{!}{\includegraphics{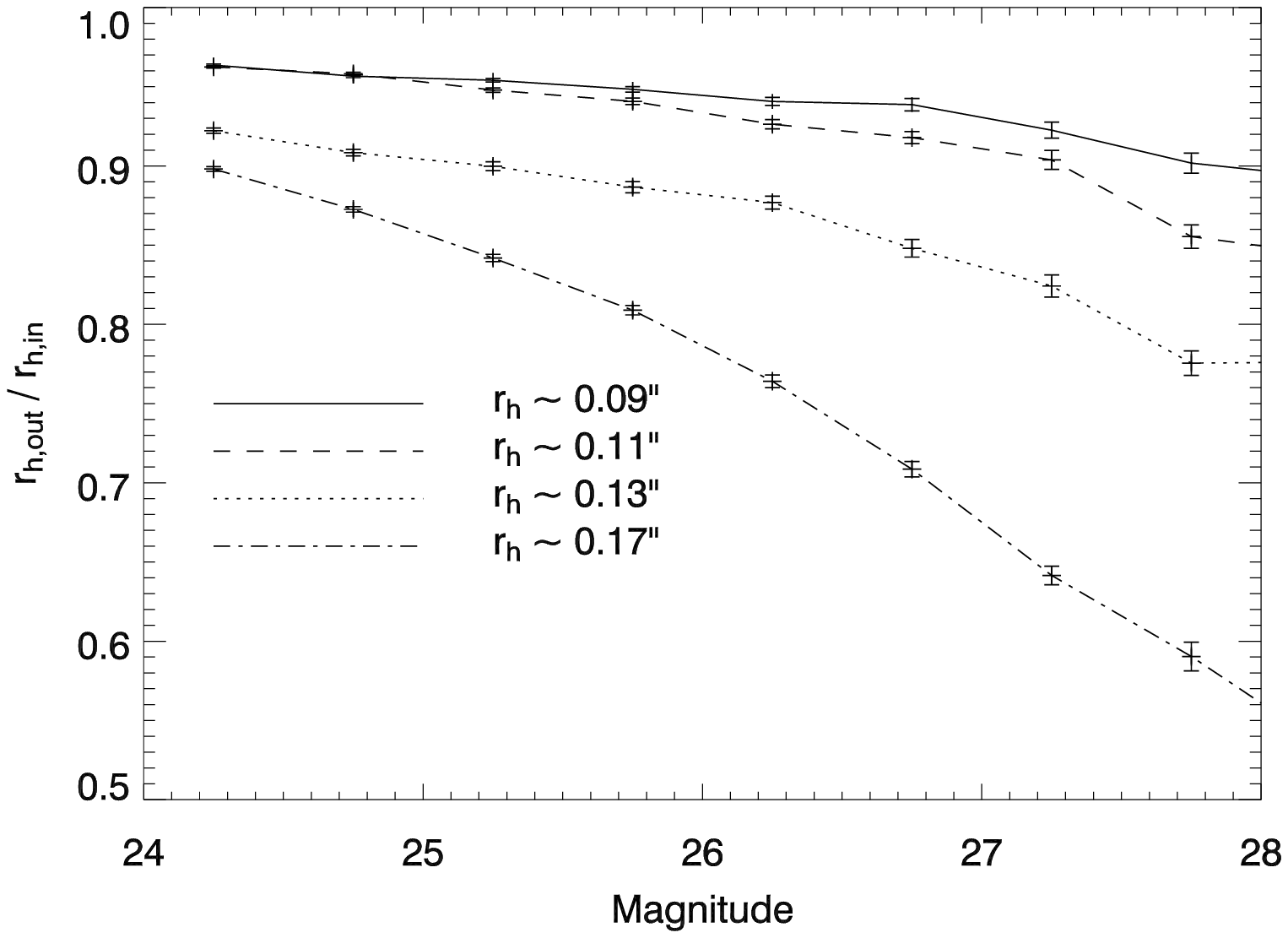}}
\caption{The ratio of recovered size over input size as a function of
  magnitude in the ACS image. The fainter and/or larger the objects,
  the more the size is underestimated.}
\label{acssizes}
\end{figure}

To quantify the size of these objects, the half light radius ($r_h$)
of each emitter was measured using the program SExtractor. The half
light radius is defined as the radius of a circular aperture in which
the flux is 50\% of the total flux. However, as already discussed in
Sect.\ \ref{photometry} and shown in Fig.\ \ref{correction}, the
fraction of the total flux of an object that is missed by SExtractor
increases when the object is fainter. As a consequence, the half light
radius that is measured by SExtractor would underestimate the size of
the object. To determine how much the half light radius was
underestimated, galaxies with a range of sizes were varied in
brightness and added to the ACS image, and the half light radii of
those objects was measured by SExtractor. It was found, as mentioned
above, that the fainter the object, the smaller its measured $r_h$, an
effect that was stronger for larger galaxies, see Fig.\
\ref{acssizes}. Using the results of these simulations, an attempt
could be made to correct the measured sizes of the confirmed \lya\
emitters. Unfortunately, this correction could overestimate the true
size of compact objects (i.e.\ objects with a half light radius
similar to that of stars). However, this only strengthens our
conclusions (see below). In Table 5 the sizes of the
emitters in the $I_{814}$-band are printed. The half light radii of
the emitters range from 0\farcs06 to 0\farcs18. The error in the half
light radius is defined as the half light radius divided by the
signal-to-noise of the object. Translating the sizes directly to
physical sizes, the measured half light radii correspond to 0.5--1.5
kpc. The median size is $\sim$1 kpc.

The mean half light radius of isolated, unsaturated stars in
the field was found to be $\sim$0\farcs06. Four of the emitters in the
ACS field have a $r_h$ that is within 1 $\sigma$ equal to the half
light radii of the stars in the field. These four emitters are
classified as unresolved (Fig.\ \ref{unresolved}).

The sizes of the confirmed \lya\ emitters can be compared to other
high redshift galaxies, e.g.\ LBGs. Recently, sizes were measured of
galaxies at various redshifts in the Great Observatories Origins Deep
Survey \citep[GOODS,][]{fer04}. For their analysis, they used
SExtractor with circular apertures having a radius that is 10 times
larger than the first radial moment of the light distribution to
ensure that all the flux was inside the aperture. The survey was
restricted to rest-frame luminosities between 0.7 $L^*$ and 5
$L^*$. Using the luminosity function of LBGs derived by \citet{ste99},
this corresponds to a magnitude range of $22.78 < m_R < 24.92$. Only
two confirmed emitters located in the ACS field (\#1518 and \#1867)
satisfy the luminosity criterion used in the GOODS analysis. Emitter
\#1867 is resolved into several clumps of emission. The size of
emitter \#1518 measured with the same input parameter as
\citet{fer04}, is 0\farcs106 $\pm$ 0\farcs006, consistent with the
0\farcs102 $\pm$ 0\farcs003 derived using our own input
parameters. The half light radius of emitter \#1518 is among the
smallest Ferguson et al.\ are finding. The average size of LBGs at $z
\sim 3$ is 0\farcs28 ($\sim 2.3$ kpc). Thus, the \lya\ emitters are
small compared to LBGs at the same redshift, provided that the method we
used to measure the sizes of the \lya\ emitting galaxies gives
comparable half light radii as the approach of Ferguson et al.\ (as
was the case for emitter \# 1518).

\begin{table}
\caption{Half light radii of the confirmed emitters located within the
  field of the ACS.} 
\begin{tabular}{cccc}
\hline
\hline
Object & $r_h$ (\arcsec) & $r_h$ (kpc) & s/n \\ 
\hline
344 & 0\farcs07 $\pm$ 0\farcs03 & $<$ 0.8 & 4.8 \\
695 & 0\farcs10 $\pm$ 0\farcs02 & 0.6 $\pm$ 0.2 & 10.5 \\
995 & 0\farcs09 $\pm$ 0\farcs01 & 0.5 $\pm$ 0.2 & 10.2 \\
1203 & 0\farcs06 $\pm$ 0\farcs01 & $<$ 0.6 & 7.1 \\
1446 & 0\farcs14 $\pm$ 0\farcs02 & 1.0 $\pm$ 0.2 & 9.4 \\
1498 & 0\farcs18 $\pm$ 0\farcs06 & 1.4 $\pm$ 0.5 & 6.0 \\
1518 & 0\farcs10 $\pm$ 0\farcs01 & 0.7 $\pm$ 0.1 & 43.1 \\
1612 & 0\farcs14 $\pm$ 0\farcs05 & 1.0 $\pm$ 0.5 & 3.8 \\
1710 & 0\farcs10 $\pm$ 0\farcs03 & 0.6 $\pm$ 0.4 & 5.8 \\
1753 & Clumpy & -- & $\sim 12$ \\
1829 & Clumpy & -- & $\sim 7$ \\
1867 & Clumpy & -- & $\sim 20$ \\
1891 & 0\farcs10 $\pm$ 0\farcs03 & $<$ 1.3 & 5.5 \\
1955 & 0\farcs08 $\pm$ 0\farcs03 & $<$ 1.0 & 6.1 \\
1962 & 0\farcs13 $\pm$ 0\farcs01 & 1.0 $\pm$ 0.1 & 12.8 \\
1968 & 0\farcs14 $\pm$ 0\farcs02 & 1.1 $\pm$ 0.2 & 9.0 \\
HzRG & Clumpy & -- & $\sim 35$ \\
\hline
\end{tabular}
\end{table}

\subsection{Star formation rate}
\label{sfr}

The average star formation rate (SFR) of the confirmed emitters, as
derived from the \lya\ flux (see Table 2), is 2.5 \msunyr\
(excluding the radio galaxy and the QSO, emitter \#2487). This
calculation assumed a \lya /H$\alpha$ ratio of 8.7 \citep[Case B
recombination,][]{bro71} and a H$\alpha$ luminosity to SFR conversion
for a Salpeter initial mass function (IMF) from \citet{mad98}:

\begin{equation}
SFR_{\mathrm{H}\alpha} = \frac{L_{\mathrm{H}\alpha}}{1.6 \times
10^{41}~\mathrm{erg}\,\mathrm{s}^{-1}}.
\end{equation}

\noindent
Because of \lya\ absorption (see e.g.\ Fig.\
\ref{spec344995}--\ref{spec33881687}), this SFR calculation gives a
lower limit.

\begin{figure}[!t]
\resizebox{\hsize}{!}{\includegraphics{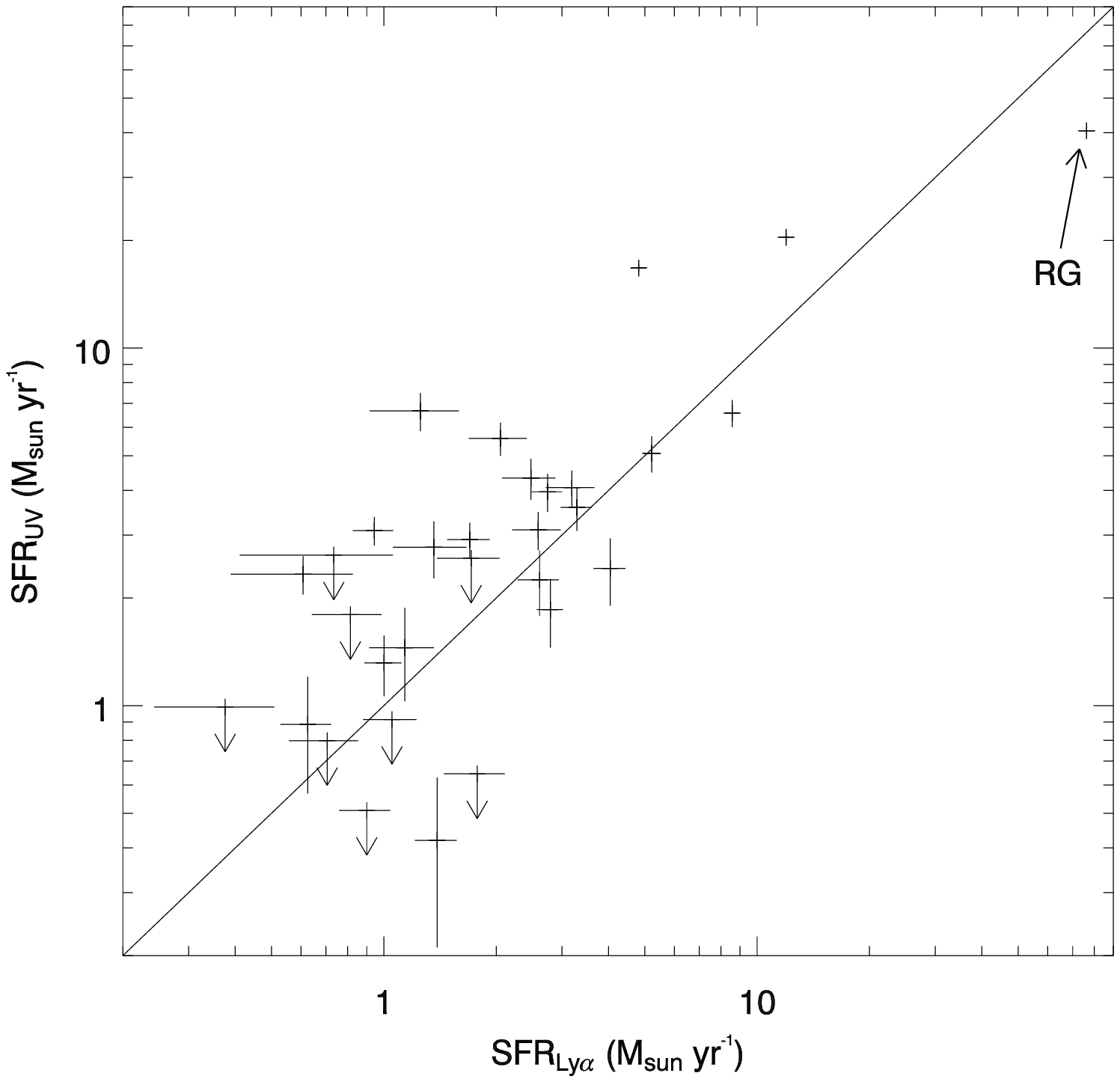}}
\caption{SFR calculated from the UV continuum plotted against the SFR
  computed using the \lya\ flux for the confirmed \lya\ emitters.}
\label{sfrplot}
\end{figure}

An alternative way to estimate the SFR is to use the level of the UV
continuum. The flux density at a wavelength of $\lambda_\mathrm{rest}
= 1500$ \AA\ can be converted to a SFR following the relation

\begin{equation}
SFR_\mathrm{UV} = \frac{L_\mathrm{UV}(\lambda_\mathrm{rest} =
1500~ \mathrm{\AA})}{8.0 \times 10^{27}~
\mathrm{erg}\,\mathrm{s}^{-1}\,\mathrm{Hz}^{-1}}
\end{equation}

\noindent
for a Salpeter IMF \citep{mad98}. In Fig.\ \ref{sfrplot} the SFR as
calculated from the \lya\ emission is plotted against the UV continuum
SFR. On average, the two methods to calculate the SFR give the same
result, with the SFR measured from the UV continuum a factor 1--1.5
higher than the \lya\ inferred SFR. The average SFR of the emitters as
measured by the UV continuum is $\lesssim$ 3.8 \msunyr. This is much
lower than the average SFR of LBGs, which is somewhere between 10 and
100 \msunyr\ \cite[e.g.][]{gia02}.

Recent measurements of the polarization of the UV continuum of the
radio galaxy indicate that the UV continuum is dominated by emission
from stars. The contribution from a scattered AGN is small, which is 
implied by the upper limit on the polarization of the continuum of $P
< 4$\% (C. De Breuck et al., in preparation). If all the light at a
rest-frame wavelength of 1500 \AA\ is due to young stars, then the SFR
of the radio galaxy is 40.5 $\pm$ 0.8 \msunyr. No correction is made
for dust absorption. This is similar to the uncorrected SFR (as
calculated from the rest-frame UV continuum) in radio galaxies at $z
\sim 2.5$ \citep[e.g.][]{ver01} and a factor of $\sim 5$ lower than
the SFR of the radio galaxy 4C41.17 at $z=3.8$ \citep{cha90,dey97}.

\section{Notes on individual objects}
\label{notes}

\begin{figure*}
\includegraphics[angle=-90,width=17cm]{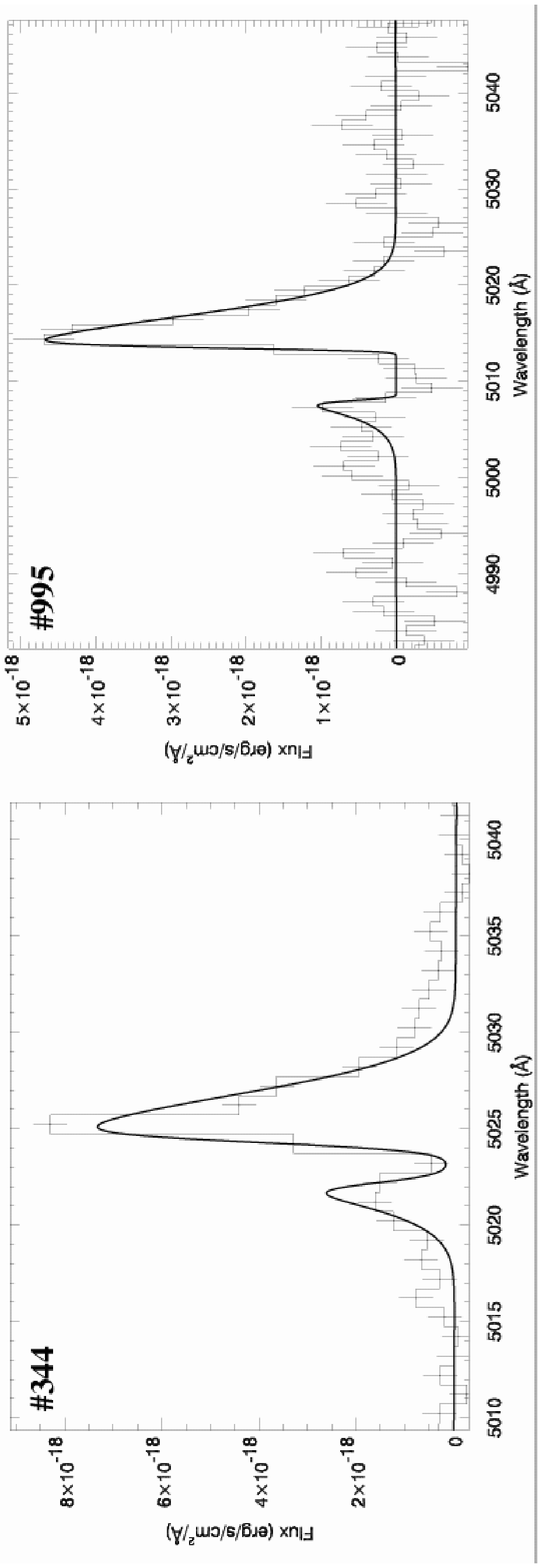}
\caption{\label{spec344995} Spectra of emitter \#344 (left) and \#995 (right).}
\end{figure*}

\begin{figure*}
\includegraphics[angle=-90,width=17cm]{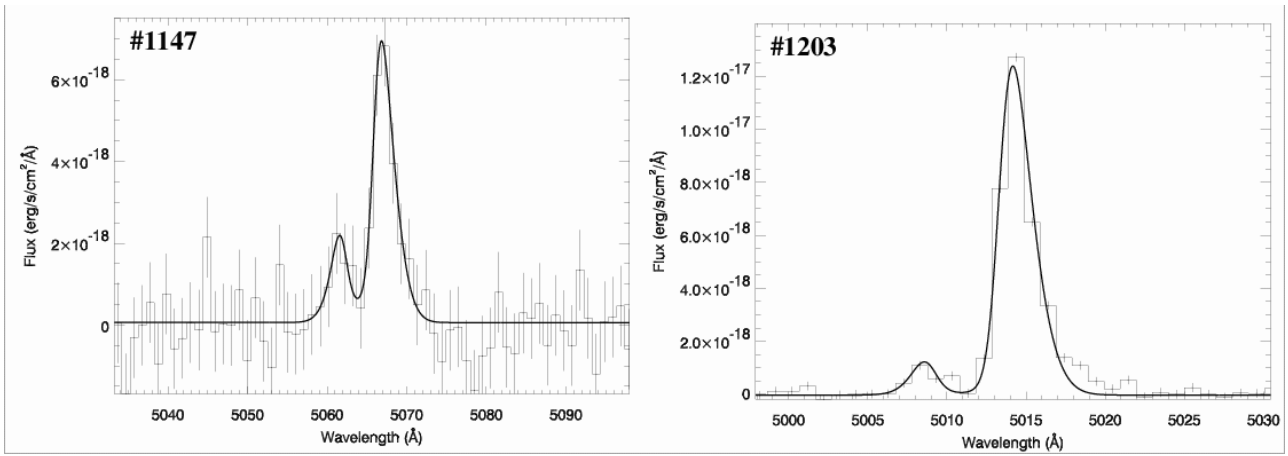}
\caption{\label{spec11471203} Spectra of emitter \#1147 (left) and \#1203 (right).}
\end{figure*}

\begin{figure*}
\includegraphics[angle=-90,width=17cm]{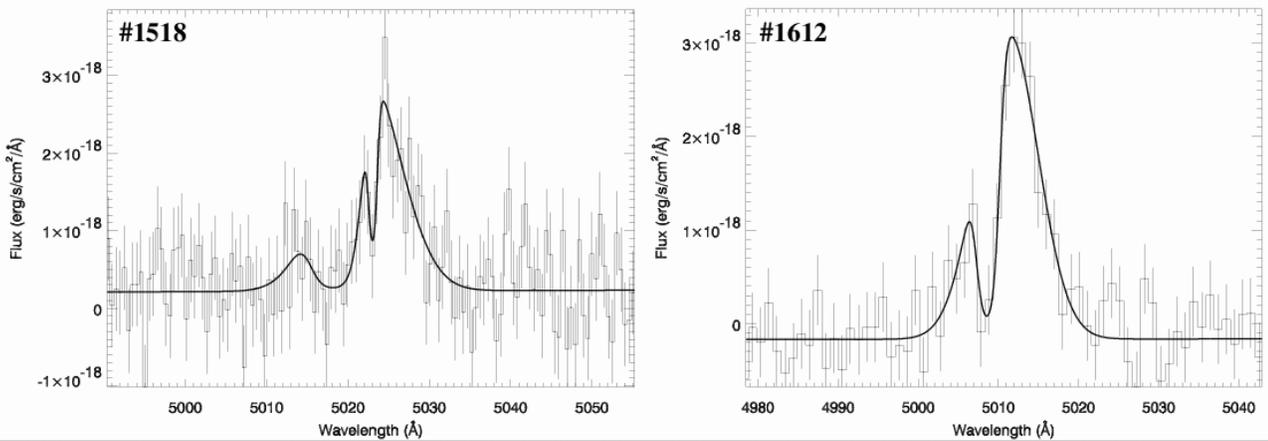}
\caption{\label{spec15181612} Spectra of emitter \#1518 (left) and \#1612 (right).}
\end{figure*}

\begin{figure*}
\includegraphics[angle=-90,width=17cm]{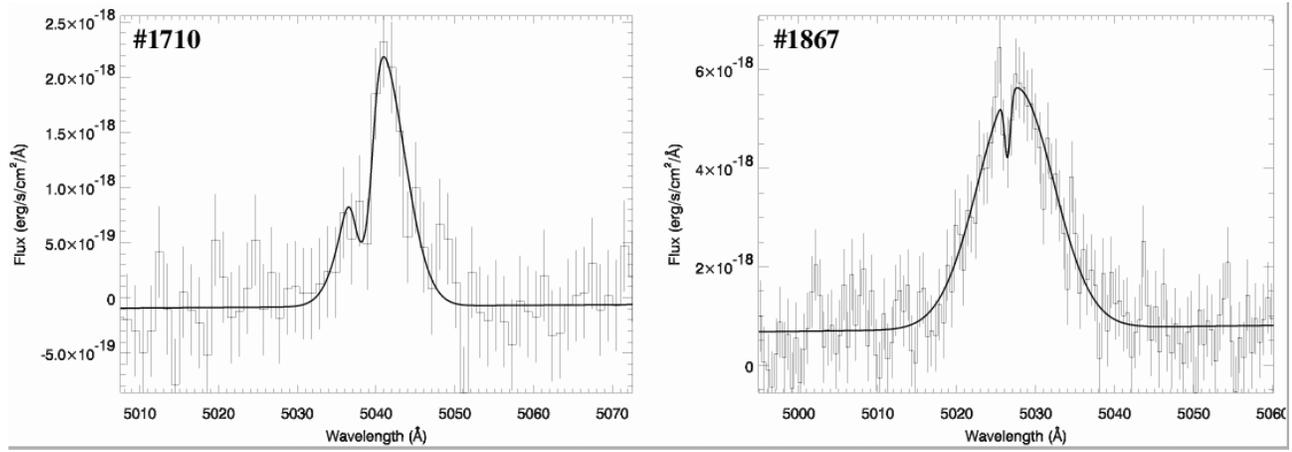}
\caption{\label{spec17101867} Spectra of emitter \#1710 (left) and \#1867 (right).}
\end{figure*}

\begin{itemize}

\item {\bf \#344}: The spectrum of this emitter is shown in Fig.\
  \ref{spec344995}. The \lya\ line can be fitted by a combination of a
  Gaussian and a Voigt absorption profile which is located $80 \pm 10$
  \kms\ blueward of the peak of the Gaussian. The fit is shown as
  the solid line in Fig.\ \ref{spec344995}. From the 2D spectrum a
  lower limit of 2\arcsec\ ($\sim 16$ kpc) on the linear size of the
  absorber could be derived, giving a lower limit
  on the \ion{H}{i} mass of $M$(\ion{H}{i}) $>$ 550 {\em M}$_{\sun}$.

\item {\bf \#995}: The spectrum of this object can be fitted by a
  Gaussian with a narrow Voigt profile centred $150 \pm 20$ \kms\ blueward
  of the emission peak (solid line in Fig.\ \ref{spec344995}). The
  column density of the absorbing neutral hydrogen is nearly
  unconstraint by our spectrum and lies in the range
  $N$(\ion{H}{i}) $\sim 10^{13.5 - 18.5}$. The absorber has a size of
  $> 2$\arcsec\ in the 2D spectrum and taking a column density of $10^{16}$
  cm$^{-2}$, the inferred mass of neutral hydrogen is at least $2
  \times 10^4$ {\em M}$_{\sun}$.

\item {\bf \#1147}: This object is one of the two emitters that were
  not detected in the ACS image down to a 3 sigma magnitude limit of
  27.1 mag\,arcsec$^{-2}$. Its redshift is $\sim 2400$ \kms\ redder than
  the median redshift of the emitters, and is therefore unlikely to be
  associated with the protocluster (see Sect.\ \ref{veldistsec}). The
  spectrum shows absorption which is located $70 \pm 20$ \kms\ blueward of the
  peak of the unabsorbed emission (Fig.\ \ref{spec11471203}). A lower limit
  of 550 {\em M}$_{\sun}$ can be given for the \ion{H}{i} mass.

\item {\bf \#1203}: This emitter is unresolved in both the VLT and ACS
  images (Fig.\ \ref{unresolved}). The spectrum is shown in Fig.\
  \ref{spec11471203}. The absorption is located $90 \pm 10$ \kms\ to
  the blue of the peak of the Gaussian. We obtain a lower limit for
  the mass of neutral hydrogen responsible for the absorption of $>$
  1700 {\em M}$_{\sun}$.

\item {\bf \#1446}: The emitter has a very blue continuum slope
  ($\beta = -4.88 \pm 0.96$). The computed equivalent width of {\em EW}$_0 =
  12$ \AA\ is below the selection criterion ({\em EW}$_0 > 15$ \AA). The VLT
  $I$-band magnitude and the $I_{814}$ magnitude derived from the ACS
  image are different at the 2.8 $\sigma$ level, taking the
  differences in effective wavelength of the filters into
  account. Using the ACS magnitude the continuum slope becomes $-2.61$
  and the {\em EW}$_0 \sim 20$ \AA. The object is resolved in the ACS image
  and has a half light radius of $1.0 \pm 0.2$ kpc (Fig.\
  \ref{resolved}). 

\item {\bf \#1498}: In the ACS image, an object with $r_h = 1.4 \pm
  0.5$ kpc is located $\sim$0\farcs5 from the position of the weak emitter
  in the VLT image.

\item {\bf \#1518}: This is the fourth brightest \lya\ emitter in the
  field. The object is extended in both the \lya\ image and the ACS
  image (see Sect.\ \ref{morphsec} and Fig.\ \ref{resolved}). The
  emission line can be fitted by a Gaussian with two Voigt profiles
  superimposed, one $80 \pm 50$ \kms\ to the red and the other $210
  \pm 60$ \kms\ to the blue of the Gaussian (Fig.\
  \ref{spec15181612}). This results in a lower limit of the mass of
  \ion{H}{i} of $2300$ {\em M}$_{\sun}$.

\item {\bf \#1612}: This emitter has a faint continuum ($I_{814} =
  27.84 \pm 0.28$) and is barely detected in ACS image
  (signal-to-noise of $\sim$4), and is marginally resolved (Table
  5, Fig.\ \ref{resolved}). The spectrum shows absorption
  of $10^{14.4 \pm 0.1}$ cm$^{-2}$ \ion{H}{i}, located $150 \pm 10$
  \kms\ to the blue of the redshift of this galaxy (Fig.\
  \ref{spec15181612}).

\item {\bf \#1710}: This is a blue emitter ($\beta = -2.26 \pm 1.40$)
  with an absorption trough on the blue wing of the emission line (see
  Fig.\ \ref{spec17101867}), the result of at least 200 {\em M}$_{\sun}$ of
  \ion{H}{i}. 

\item {\bf \#1753}: At the position of this \lya\ emitter, the ACS image
  shows three separate objects located within $\sim 8$ kpc (Fig.\
  \ref{multiple}). 

\item {\bf \#1829}: This object is resolved by the ACS into an elongated
  structure consisting of several objects (Fig.\ \ref{multiple}). 

\item {\bf \#1867}: Denoted as galaxy ``A'' by LF96, a spectrum of
  this object taken under bad seeing conditions confirms the redshift
  measured by LF96 (Table 2). The \lya\ line is asymmetric
  and can be fitted by a Gaussian and one Voigt absorber, which is
  $60 \pm 20$ \kms\ blueward of the \lya\ peak (Fig.\
  \ref{spec17101867}). The VLT narrow-band image shows an extended
  \lya\ halo of $\sim$25 kpc (Fig.\ \ref{multiple}), while in ACS
  image the object is very clumpy.

\item {\bf \#1968}: This emitter was undetected in the VLT $I$-band,
  but in the ACS image an object with a half light radius of $1.1 \pm
  0.2$ kpc is visible.
 
\item {\bf \#2487}: This emitter has the brightest \lya\ line in the
  field after the radio galaxy, and is called galaxy ``B'' in LF96. As
  mentioned in Sect.\ \ref{spectra}, the \lya\ line is broad ({\em FWHM}
  $\sim$ 2500 \kms) which is most likely caused by an AGN. The
  spectrum is characterized by a large absorption trough with a column
  density of $N$(\ion{H}{i})$\sim 10^{14.6}$ cm$^{-2}$. Furthermore, the
  red wing of the \lya\ line is much broader than the blue wing. The
  spectrum can be fitted with two absorbers, located $250 \pm 170$ \kms\
  to the red and $1150 \pm 200$ \kms\ to the blue of the centre of the
  emission line. The inferred mass of \ion{H}{i} is $> 5 \times 10^4$
  {\em M}$_{\sun}$. 

\item {\bf \#2719}: This is galaxy ``C'' from LF96. Galaxy ``C'' was
  not selected as a candidate \lya\ emitter in our images. It has
  colors comparable to those quoted in LF96, but an {\em EW}$_0$ of
  $1.0^{+1.2}_{-1.1}$. LF96 measured an {\em EW}$_0$ $> 12$ \AA\ and a line
  flux of $\sim 5 \times 10^{-17}$ \ergscm, although no spectrum was
  taken of this object to confirm the existence of the line. An
  explanation for the fact that this galaxy is not selected by us as
  an emission line candidate could be that the large width of the
  narrow-band filter used by LF96, making it sensitive to a wider
  redshift range than our filter. Their filter was sensitive to \lya\
  emitters having redshifts in the range $z = 3.08 - 3.16$, while our
  filter is sensitive to the redshift range $z = 3.12 - 3.17$. Galaxy
  ``C'' could be a \lya\ emitter with a redshift between $z = 3.08$
  and $z=3.12$, and it would therefore be part of the protocluster,
  but not be included as one of our candidates (see Sect.\
  \ref{veldistsec}).

\item {\bf \#3101}: The \lya\ line of this emitter is broad ($800 \pm
  100$ \kms\ {\em FWHM}, see Fig.\ \ref{spec24873101}), as compared to the
  median line width of the emitters (260 \kms). The spectrum can be
  fitted by a Gaussian, superimposed by two Voigt absorbers located $40
  \pm 110$ and $240 \pm 20$ \kms\ to the blue of the emission, the
  result of at least $\sim 1000$ {\em M}$_{\sun}$ of \ion{H}{i}.

\item {\bf \#3388}: This blue \lya\ emitter ($\beta = -1.92 \pm 0.52$)
  shows an absorption trough $130 \pm 10$ \kms\ blueward of the
  emission redshift (Fig.\ \ref{spec33881687}), implying a neutral
  Hydrogen mass of $> 625$ {\em M}$_{\sun}$.

\item {\bf Radio galaxy MRC 0316--257}: The absorption structure of
  the radio galaxy is complicated. Only a Gaussian emission line
  profile with 4 separate absorbers gives a reasonable fit (solid line
  in Fig.\ \ref{spec33881687} and Table 3). The absorbers
  are $200 \pm 10$ \kms\ to the red of the peak of the Gaussian and
  $270 \pm 10$ \kms, $660 \pm 10$ \kms\ and $970 \pm 20$ \kms\ to the
  blue. Approximately 2\arcsec\ to the north-east of the radio galaxy
  is a foreground galaxy, with a clear spiral structure in the ACS
  image. An emission line was detected in a spectrum of this object,
  with a wavelength around 6965 \AA, most likely [\ion{O}{ii}] at $z \sim
  0.87$ (C. De Breuck, private communications).

\end{itemize}

\section{A protocluster at $z=3.13$?} 
\label{protocluster}

\subsection{Volume density}
\label{dgal}

Several surveys have been carried out to estimate the (field) volume density
of \lya\ emitters at $z \sim 3$
\citep[e.g.][]{cow98,hu98,kud00,cia02,fyn03}. To estimate the
(over)density of \lya\ emitters in our field, we will compare our
numbers with those found in the survey of \citet{cia02}, since it
covers the largest area of all the surveys, and of \citet{fyn03},
because they present the deepest and spectroscopically most complete
comparison sample of blank field $z \sim 3$ \lya\ emitters.

\citet{cia02} made a blank field survey to estimate the density of
emission line sources and to calculate the contamination of
intra-cluster planetary nebulae searches. They searched for faint
emission line sources in a 0.13 deg$^2$ field at a wavelength of 5019
\AA. They found 21 objects with an observed equivalent width greater
than 82 \AA\ and a $m_\mathrm{nb} < 24.3$. Assuming all their sources
are \lya\ emitters at $z\sim3.13$, the volume density of field
emitters is $n_{\mathrm{field}} = 2.4^{+0.6}_{-0.5} \times 10^{-4}$
Mpc$^{-3}$. Applying the same selection criteria to our data, 5
emission line objects (excluding the radio galaxy) are found in the
field around 0316--257, all confirmed to be \lya\ emitters at $z\sim
3.13$. This gives a density of $n_{\mathrm{0316}} = 5.4^{+3.7}_{-2.3} 
\times 10^{-4}$ Mpc$^{-3}$. The density in the 0316 field is therefore
a factor $2.2^{+1.8}_{-1.0}$ higher than the field density. The large errors
on this number are due to small number statistics. 

More recently, \citet{fyn03} presented the first results of a program
to detect faint \lya\ emitters at $z \sim 3$. They used the same VLT
narrow-band filter as described in Sect.\ \ref{ima} to image a field
that contained a damped \lya\ absorber. The 5$\sigma$ detection limit
for point sources in their narrow-band image is $m_\mathrm{nb} < 26.5$
as measured in a circular aperture with a size twice the seeing {\em FWHM},
is very similar to our 5$\sigma$ detection limit ($m_\mathrm{nb} <
26.4$). They found 27 candidate \lya\ emitters with an equivalent
width greater than 12.5--25 \AA, the limit depending on the predicted
line flux. Subsequent spectroscopy confirmed that 18 of the 22
candidate emitters observed are \lya\ emitters and two were foreground
[\ion{O}{ii}] emitters. Assuming that the seven unconfirmed candidate
emitters are all \lya\ emitters at $z \sim 3$, the number of \lya\
emitters down to a flux limit of $6 \times 10^{-18}$ \ergscm\ in their
field is 25. Our number of emitters selected with the same equivalent
width limits is $\sim$75 after correction for foreground
contaminants. This implies a density of $3.0^{+0.9}_{-0.7}$ times the
field density. Roughly, we find three times the number of \lya\
emitters as might be expected from field surveys.

\citet{mai03} gathered measured abundances of \lya\ emitters from the
literature, shifted them to $z=3.5$ and fitted a model function
through the points \citep[a description of the model can be found
in][]{tho04}. They predict approximately 2325 \lya\ emitters per deg$^2$ in
a volume with $\Delta z = 0.1$ with line fluxes exceeding $5 \times
10^{-18}$ \ergscm. If their model is correct, then $\sim 15$ \lya\
emitting galaxies at $z = 3.13$ are expected within our volume
brighter than $7 \times 10^{-18}$ \ergscm. Applying this limit, we
find 63 galaxies or 59 galaxies if we correct for possible foreground
contaminants. Of these, 29 are spectroscopically confirmed. The
density is a factor $4.0^{+0.6}_{-0.5}$ higher than the model prediction, in
agreement with the above estimates.

To summarize, the density of \lya\ emitters near the radio galaxy is a
factor 2--4 higher than the field density, indicating that the radio
galaxy might reside in an overdense region.

\subsection{Velocity distribution}
\label{veldistsec}

\begin{figure}[!t]
\resizebox{\hsize}{!}{\includegraphics{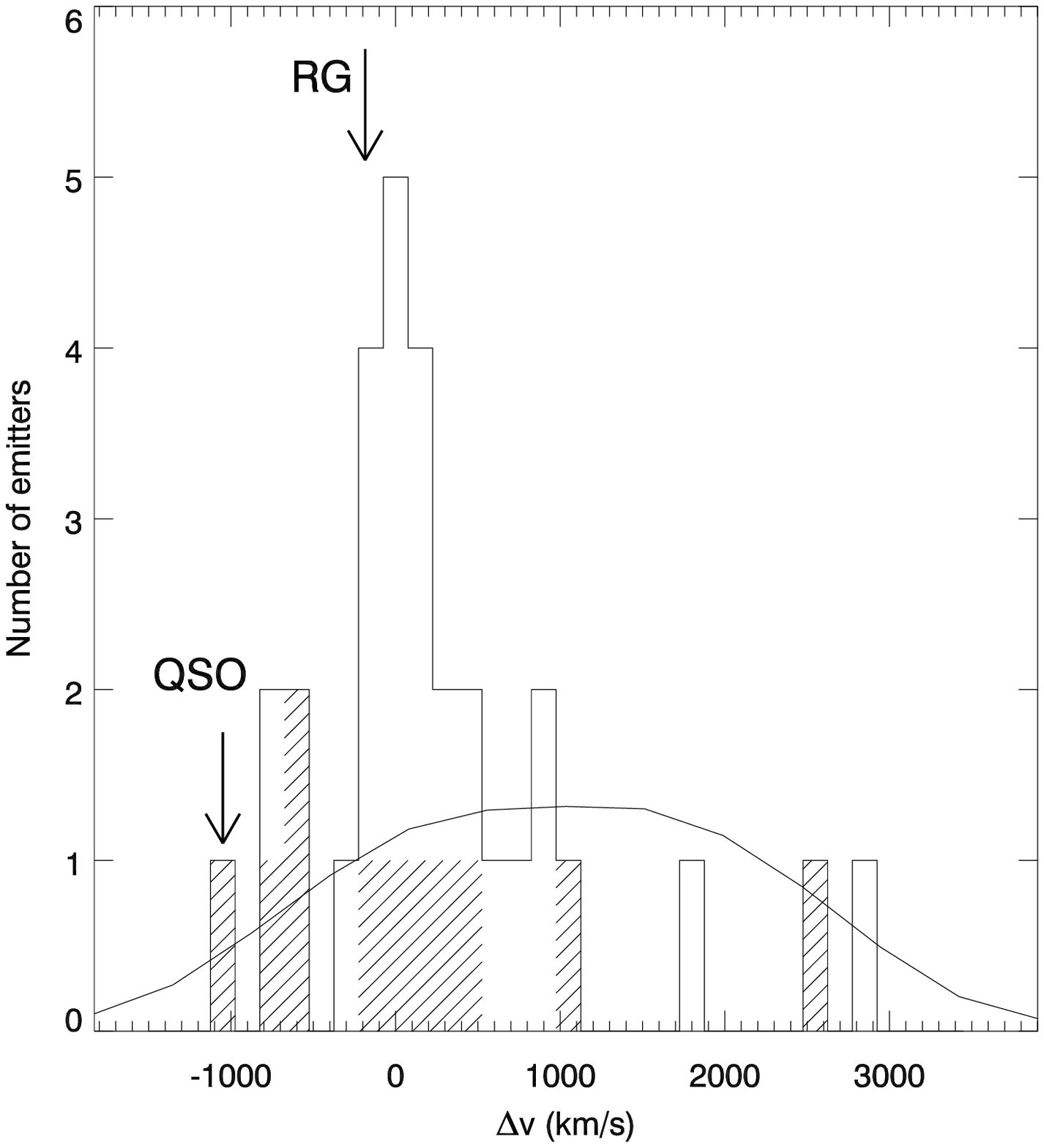}}
\caption{Velocity distribution of the confirmed \lya\ emitters. The
  bin size is 150 \kms. The median redshift of the confirmed emitters
  $z = 3.1313$ is taken as the zero-point. The velocities of the radio
  galaxy and the QSO are indicated with arrows. The solid line
  represents the selection function of the narrow-band filter,
  normalized to the total number of emitters found. The hashed
  histogram represents the velocities of emitters with absorption in
  their line profile. The three objects in the red part of the filter
  (with a velocity greater than 1500 \kms) might not be part of the
  protocluster.}
\label{veldist}
\end{figure}

The velocity distribution of the emitters is plotted in Fig.\
\ref{veldist}. The response of the narrow-band filter used to select
the candidate emitters for spectroscopy is also shown. Interestingly,
the emitters are not distributed homogeneously over the filter, but
most of them appear to cluster on the blue side of the filter. The
emitters which show absorption in their emission line profiles seem to
be distributed more homogeneously, but this could be due to the small
number of objects. 

To test whether the clustering of emitters in redshift space is
significant, Monte Carlo simulations of the redshift distribution were
performed. 10000 realizations with 31 emitters were reproduced, with
the narrow-band filter curve as the redshift probability function for
each emitter. The mean of the observed redshift distribution
($z=3.136$) differs 2.6 $\sigma$ compared to the simulated
distribution ($z=3.146 \pm 0.004$) and the width of the observed
redshift distribution differs by 1.7 $\sigma$ (0.012 compared to 0.022
$\pm$ 0.006). In total, the measured redshift distribution deviates
from the simulated one by 3.07 $\sigma$. This means that a redshift
distribution as observed was reproduced in only 0.2\% of the
cases. The peak of velocity distribution of the \lya\ emitters lies
within 200 \kms\ of the redshift of the radio galaxy (Fig.\
\ref{veldist}), providing evidence that most of the \lya\ emitters are
physically associated with the radio galaxy. Taking together, the
observed overdensity of \lya\ emitters in our field (Sect.\
\ref{dgal}), combined with the peak in the redshift distribution
provide compelling evidence that the \lya\ emitters reside in a
protocluster at $z \sim 3.13$.

\section{Properties of the protocluster}
\label{protoprop}

In this Section, we discuss the macro properties of the \lya\ emitters
in the protocluster. 

\subsection{Velocity dispersion}
\label{veldisp}

To determine the velocity dispersion of the emitters, the biweight
scale estimator was used \citep{bee90}. This is the most appropriate
scale estimator for samples of 20--50 objects \citep{bee90}. The
velocity dispersion is 640 $\pm$ 195 \kms, corresponding to a {\em
FWHM} of 1510 $\pm$ 460 \kms. This is significantly smaller than the
width of the narrow-band filter, which has a {\em FWHM} of $\sim 3500$
\kms. Although most \lya\ emitters are likely to be members of the
protocluster, the three emitters with velocities $>$ 1500 \kms\ from
the peak of the distribution (Fig.\ \ref{veldist}) are probably field
galaxies. Ignoring these three field galaxies, the velocity dispersion
drops to $535 \pm 100$ \kms. On the lower redshift side (negative
velocities), no clear edge is visible in the distribution. This could
be due to the decrease in sensitivity of the narrow-band filter on this
side of the redshift distribution. If the protocluster extends to
much lower redshifts, our estimate of the velocity dispersion is
a lower limit.

\subsection{Spatial distribution}
\label{spadist}

\begin{figure}[!ht]
\resizebox{\hsize}{!}{\includegraphics{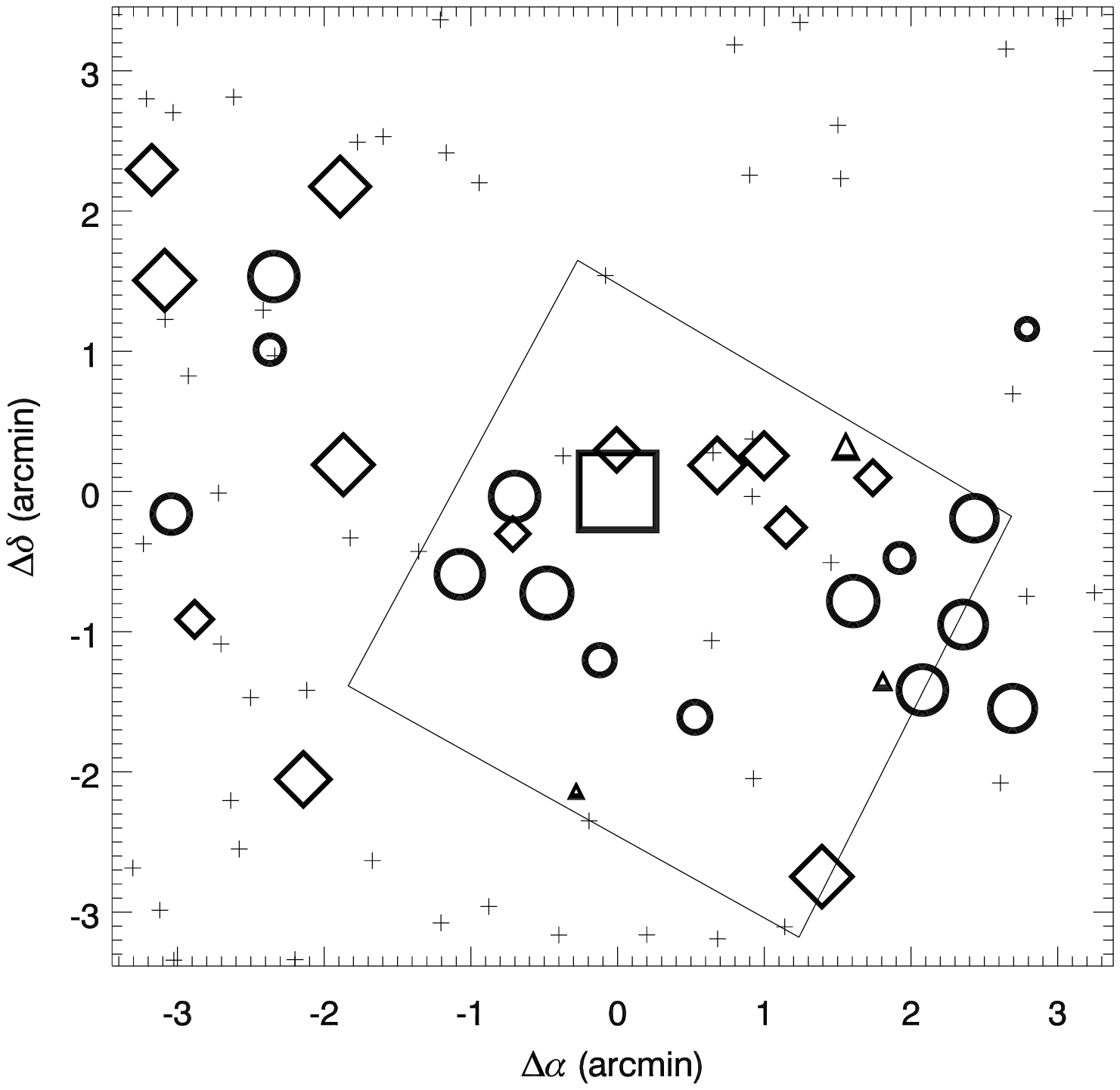}}
\caption{Spatial distribution of \lya\ emitters around the radio
galaxy MRC 0316--257 (denoted by a square). The confirmed emitters in
the protocluster are represented by circles (emitters with a redshift
smaller than the median, $z < 3.1313$) and diamonds (emitters with $z
> 3.1313$). The triangles show the position of the three \lya\
emitting galaxies with a velocity $> 1500$ \kms\ from the median
velocity of the emitters. The sizes of the symbols are scaled according
to the velocity offset from the median, with larger
symbols representing emitters with a redshift closer to the median
redshift. The pluses are objects satisfying the selection criteria
(see Sect.\ \ref{selection}), but are not (yet) confirmed. The
quadrangle represents the outline of the ACS image.}
\label{skydist}
\end{figure}

The spatial distribution of the emitters is shown in Fig.\
\ref{skydist}, where the circles, diamonds and triangles represent the
spectroscopically confirmed emitters, and with the sizes of the
symbols scaled according to the velocity offset from the median of the
emitters. The pluses are unconfirmed candidate \lya\ emitters
satisfying our selection criteria (see Sect.\ \ref{selection}). The
majority of these candidates (96\%) has not yet been observed
spectroscopically, while the remaining 4\% were too faint to be
confirmed. The imaging field of view (3.3 $\times$ 3.3 Mpc$^2$ at
$z=3.13$) is not large enough to show clear boundaries of the
structure.

\subsection{Mass}
\label{mass}

At a redshift of $z=3.13$, the age of the Universe is only 2.2
Gyr. Taking the velocity dispersion as a typical velocity for a galaxy
in the protocluster, it would take at least 5 Gyr to cross the
structure, making it highly unlikely that the protocluster is near
virialization. Therefore, the virial theorem cannot be used to
calculate the mass of the protocluster. 

Another way to compute the mass is to use the (comoving) volume $V$
occupied by the overdensity, the (current) mean density of the
Universe $\bar{\rho}$ and the mass overdensity of the protocluster
$\delta_m$:

\begin{equation}
M=\bar{\rho}\,V\,(1 + \delta_m) = \bar{\rho}\,V\,(1 +
\delta_{\mathrm{gal}}/b).
\end{equation}

\noindent
where $b$ is the bias parameter ($b \equiv
\delta_{\mathrm{gal}}/\delta_m$), relating the observed galaxy
overdensity ($\delta_{\mathrm{gal}} = n_{0316}/n_{\mathrm{field}} -
1$) to the mass overdensity and $\bar{\rho} = 3.5 \times 10^{10}$ {\em
M}$_{\sun}$\,Mpc$^{-3}$ for the cosmological parameters used in this
paper.

The weighted mean of the three density estimates in Sect.\ \ref{dgal}
is $n_{0316}/n_{\mathrm{field}} = 3.3^{+0.5}_{-0.4}$, giving an
overdensity of $\delta_{\mathrm{gal}} = 2.3^{+0.5}_{-0.4}$. Taking
$V = 9.3 \times 10^3$ Mpc$^3$ (Sect.\ \ref{vltdatared}) and $b = 3 - 6$
\citep{ste98,shi03} results in a mass for the protocluster within the
observed volume of $4-6 \times 10^{14}$ {\em M}$_{\sun}$. Because the size
of the protocluster is unconstrained (e.g.\ Fig.\ \ref{skydist}), this
mass estimate is a lower limit.

However, the redshift range of protocluster galaxies is likely to be
smaller than the redshifts for which the narrow-band filter is
sensitive (see Fig.\ \ref{veldist}). Assuming that the three outlying
galaxies on the red side of the filter as field galaxies, the redshift
range of the protocluster galaxies is 0.029 and the volume occupied by
these emitters is $5.4 \times 10^3$ Mpc$^3$. This estimate of the
volume does not take into account the redshift space distortions
caused by peculiar velocities \citep[][see below]{ste98}. Assuming
that in total $\sim 10$\% of the (candidate) emitters are field
galaxies (see Sect.\ \ref{veldistsec}), the density of emitters within
this volume with respect to the field density is
$1+\delta_{\mathrm{gal}} = 3.3^{+0.5}_{-0.4} \times 0.9 \times
\frac{9331}{5400} = 5.1^{+0.8}_{-0.6}$. In this approach, the relation
between the mass overdensity $\delta_m$ and the observed galaxy
overdensity $\delta_{\mathrm{gal}}$ is \citep{ste98}:

\begin{equation}
1 + b \delta_m = C(1 + \delta_{\mathrm{gal}}), 
\end{equation}

\noindent
where $C$ takes into account the redshift space distortions
\citep{ste98}. Assuming that the structure is just breaking away from
the Hubble flow, $C$ can be approximated by

\begin{equation}
C = 1 + f - f(1 + \delta_m)^{1/3}
\end{equation}

\noindent
\citep{ste98}, with $f$ the rate of growth of perturbations at the
redshift of the protocluster \citep{lah91}. $f$ not only depends on
$z$, but also on $\Omega_\mathrm{M}$ and $\Omega_{\Lambda}$. In the
cosmology adopted in this paper ($\Omega_\mathrm{M} = 0.3$ and
$\Omega_{\Lambda} = 0.7$), $f$ is close to 1 at high redshift
\citep[$z > 2$,][]{lah91}.

Again taking $b = 3 - 6$, the computed mass of the protocluster is $>
3-5 \times 10^{14}$ {\em M}$_{\sun}$. As both the redshift range of
the galaxies and the size of the protocluster could be larger than is
observed (see Sects.\ \ref{veldisp} and \ref{spadist}), this estimate is
again a lower limit. The computed mass roughly corresponds to the
virial mass of the Virgo cluster \citep[e.g.][]{fou01}.

\section{Nature of the \lya\ emitters}
\label{nature}

What can we deduce about the nature of the \lya\ emitters in our
field?  Four of the 16 emitters (25\%) detected in the ACS image are
unresolved and may be narrow-line \citep[{\em FWHM} $\lesssim$
1000--2000 \kms, e.g.][]{ben02} QSOs. The number density of faint ($21
< m_R < 25.5$) QSOs near $z = 3$ in the LBG survey of Steidel and
collaborators \citep{ste03} is 250 QSO deg$^{-2}$ in a redshift
interval $\Delta z \simeq 0.6$ \citep{ste02}. They observe a ratio of
narrow-line to broad-line QSOs of $N$(narrow)/$N$(broad) $= 1.2
\pm 0.4$ \citep{ste02}. Assuming an overdensity of galaxies in our
field of $n_{0316}/n_{\mathrm{field}} = 4.0$ (Sect.\ \ref{dgal}), the
predicted number of QSOs in field is $\sim 1$. Indeed, one
(broad-line) QSO is found (emitter \#2487). Extrapolating the faint
end of the QSO luminosity function given by \citet{hun04} to $m_R >
25.5$, the number of QSOs near MRC 0316--257 with $25.5 < m_R <
28.5$ is calculated to be $< 1$, making the identification of the
\lya\ emitters with QSO unlikely. Another reason against the
classification of the \lya\ emitters as QSOs is the continuum slope of
narrow-line QSOs in the LBG survey of $\beta = -0.4$
\citep{ste02}. This is much redder than the median of the unresolved
emitters ($\beta \sim -1.43$). We therefore conclude that all the
\lya\ emitters are star forming galaxies\footnote{We can not
entirely exclude that some of the emitters harbour a (concealed) AGN. A
40 ks exposure by the {\em Chandra} X-ray observatory is scheduled for
early 2005, which should help to estimate the fraction of the \lya\
emitters that contains an AGN.}

The next question is how the properties of the \lya\ emitters compare
to those of the LBG population as a whole.

\begin{itemize}

\item[$\bullet$] {\em Continuum luminosity:} Besides that of the radio galaxy
and the broad-line QSO, the continuum of the emitters is faint, the
brightest emitter having $m_R =$ 24.24 (using $m_R = -2.5\,
^\mathrm{10}\log(f_\nu (\lambda_\mathrm{rest} = 1700 \mathrm{\AA})) -
48.59$) and the faintest $m_R > 28.45$. Using $m_* = 24.53$ at $z =
3.13$ \citep{ste99}, this corresponds to luminosities ranging from 1.3
$L^*$ to $<$ 0.03 $L^*$. Roughly 90\% of the emitters are fainter than
$m_R =$ 25.5, the spectroscopic limit for LBGs. A similar percentage
was found by \citet{fyn03}.

\item[$\bullet$] {\em Size:} A comparison of the sizes of \lya\
emitters with those of LBGs at $z \sim 3$ suggests that \lya\ emitters
are generally smaller (with $r_h <$ 1.5 kpc) than LBGs which have an
average size of $r_h \sim$ 2.3 kpc.

\item[$\bullet$] {\em Color:} The colors of the confirmed emitters
are, on average, very blue. The median UV continuum slope is $\beta =
-1.76$, bluer than the average slope of LBGs with \lya\ emission
\citep[$\beta \sim -1.09$;][]{sha03}. A large fraction of the
confirmed emitters ($\sim 2/3$) have colors consistent with negligible
absorption and could be dust-free starburst galaxies.

\end{itemize}

\noindent
Summarizing these properties, the \lya\ emitters are on average
fainter, bluer and smaller than $z \sim 3$ LBGs.

Various models have been proposed to explain the properties of \lya\
emitters. Based on rest-frame optical photometry of LBGs,
\citet{sha01} concluded that LBGs with \lya\ in emission are ``old''
(age $>$ few $\times 10^8$ yr), while ``young'' (age $< 10^8$ yr) LBGs
have \lya\ in absorption. This could be explained by the young
galaxies being dusty which caused the absorption of \lya\ photons,
while the older galaxies are more quiescent with less dust and
superwinds which allow \lya\ photons to escape. Other groups have
suggested that strong \lya\ emitters are young star forming galaxies
which was derived from the blue colors and high equivalent widths of
the \lya\ emitters \citep[e.g.\ LF96,][]{rho01,kee02,mal02,tap04}.
Trying to connect these observations, various authors
\citep[e.g.][]{fri99,tho04} discuss a model in which a forming galaxy
has two \lya\ bright phases: an initial phase when the galaxy has
started the very first period of star formation and is still nearly
dust free. Due to supernova explosions, the interstellar medium will
be enriched with metals, and dust will form in the galaxy. This dust
will absorb the \lya\ photons, extinguishing the \lya\ emission
line. A second \lya\ bright phase occurs at a later time when strong
galactic winds facilitate the escape of \lya\ emission. 

The observations of \lya\ emitters in our field support this second
picture, and the \lya\ emitting galaxies might be young star forming
galaxies in their first starburst phase. This should be confirmed by
deep infrared observations. Modelling the spectral energy distribution
of the \lya\ emitters from the UV to the rest-frame optical should
allow the discrimination of \lya\ emitters being either young
dust-free galaxies or more evolved star forming galaxies with an
underlying old stellar population.

\section{Implications of a protocluster at $z=3.13$}
\label{implications}

\subsection{Star formation rate density}

The total UV star formation rate density (SFRD) of the confirmed emitters
(excluding the radio galaxy) within the volume of the narrow-band
filter is $\lesssim$ 0.0127 $\pm$ 0.0003 \msunyrmpc. The SFRD derived
from observations of LBGs at $z \sim 3$ with $m_R \lesssim 27$ is 0.0184
$\pm$ 0.0034 \msunyrmpc\ \citep{ste99}. Using the same magnitude
limit, we find a SFRD of 0.0109 $\pm$ 0.0002 \msunyrmpc. This is a
lower limit, because it does not include a contribution from the radio
galaxy, the emitters with only a limit on their UV SFR are ignored and
no correction has been made for incompleteness. Assuming that only
20--25\% of the star forming, UV bright galaxies at $z\sim3$ have a
\lya\ line satisfying our selection criteria \citep{ste00,sha03}, then
the SFRD around MRC 0316--257 is roughly $>$ 2.4--3.0 times higher
than in the field, in agreement with the number density of the \lya\
emitters. 

It should be noted that the total SFRD in the protocluster might be
much higher. From rest-frame UV and optical colors, \citet{ste99} and
\citet{sha01} found that the UV continuum is on average attenuated by
a factor of $\sim 5$. Also, very dusty, obscured galaxies could be
missing. For example, \citet{deb04} found an overdensity of bright
sources at 1.2 mm wavelength in the field of the protocluster near the
radio galaxy TN J1338--1942 at $z=4.1$. These objects could contribute
substantially to the total SFR within the protocluster.

\subsection{Enrichment of the intracluster medium}

It is interesting to estimate the metal production in the
protocluster surrounding MRC 0316--257. At redshift $z \sim 0.3$,
the intracluster medium (ICM) in clusters has a metallicity of
0.2--0.3 $Z_{\sun}$ \citep[e.g.][]{mus97}, showing little
evolution up to $z \sim 1.2$ \citep{toz03,has04,mau04,ros04}.

The extinction corrected total star formation rate density of
UV-bright star forming galaxies at $z \sim 3.1$ is $\sim$0.13
\msunyrmpc\ \citep{ste99,gia04}. The protocluster has a volume of $5.4
\times 10^3$ Mpc$^3$ and a galaxy overdensity of
$n_{0316}/n_{\mathrm{field}} = 5.1$ (Sect.\ \ref{mass}). This gives a
total SFR in the protocluster of $\sim$3580 \msunyr, ignoring any
contribution to the SFR from very dusty, obscured starforming
galaxies. Taking an average yield of 0.02 \citep{lia02}, this means
that $\sim$72 {\em M}$_{\sun}$ of metals are produced every year. Taking $4
\times 10^{14}$ {\em M}$_{\sun}$ as the mass of the protocluster (Sect.\
\ref{mass}) and assuming a baryon fraction of
$\Omega_b/\Omega_{\mathrm{M}} = 0.17$ \citep[e.g.][]{spe03} and
assuming that the star formation rate in the protocluster is constant
with time, then enough metals can be produced to enrich the baryons in
the protocluster to 0.2 $Z_{\sun}$ at $z \sim 1$. However, a large
fraction ($> 90$\%) of the produced metals must escape the galaxies in
which they are formed. A possible mechanism to inject the metals into
the ICM are supernova-driven outflows \citep[e.g.][]{hec95,agu01},
which are frequently seen in $z \sim 3$ LBGs
\citep[e.g.][]{pet01,ade03}. This simple calculation shows that the star
formation rate in the protocluster is high enough to enrich the ICM to
the observed value at lower redshifts.

\subsection{High redshift protoclusters associated with radio galaxies}

Based on the high volume density of \lya\ emitters near MRC 0316--257,
which is a factor of $3.3_{-0.4}^{+0.5}$ higher as compared to blank
fields, and the small velocity distribution of the confirmed emitters
({\em FWHM} of 1510 \kms) compared to the width of the narrow-band filter
({\em FWHM} $\sim 3500$ \kms), we conclude that the \lya\ emitters are
located in a protocluster of galaxies with an estimated mass of $>$
3--6 $\times 10^{14}$ {\em M}$_{\sun}$. A likely scenario is that this
protocluster will evolve and form a massive cluster of galaxies. The
radio galaxy at the centre of the protocluster has the properties
expected of the progenitor of a massive cD elliptical. The clumpy
appearance of the radio galaxy in the ACS image could be explained as
a merger of smaller subunits, and is very similar to {\em HST}
observations of other $z \sim 2 - 3$ radio galaxies
\citep[e.g.][]{pen99}.

Based on the observations obtained in our VLT large program, the
protocluster around MRC 0316--257 is not unique. We have found galaxy
overdensities around all five radio galaxies (with redshifts between
$z=2.16$ and $z=4.1$) that were studied to a similar depth as the 0316
field \citep{kur00,pen00a,ven02,ven03,kur04}. Each of these radio
galaxy field has at least 20 confirmed protocluster members, the
velocity dispersions of the protoclusters range from 300 \kms\ to 1000
\kms\ and the associated masses are $> 10^{14}$ {\em M}$_{\sun}$
\citep{ven03}. At an even higher redshift, $z=5.2$, we found a similar
overdensity of \lya\ emitters near the radio galaxy TN J0924--2201
\citep{ven04}. In a future paper, we will describe and compare the
properties of these radio galaxy protoclusters (Venemans et al.\ in
prep).

\begin{acknowledgements}
We are grateful to the staff of Paranal, Chile, for their excellent
support. We thank Michiel Reuland, Rogier Windhorst
and Andrew Zirm for useful discussions and the referee, Paul Francis,
for helpful suggestions. GKM acknowledges funding by an
Academy Professorship of the Royal Netherlands Academy of Arts and
Sciences (KNAW). The work by WvB was performed under the auspices of
the U.S.\ Department of Energy, National Nuclear Security
Administration by the University of California, Lawrence Livermore
National Laboratory under contract No.\ W-7405-Eng-48. The NRAO is
operated by associated universities Inc, under cooperative agreement
with the NSF. This work was supported by the European Community
Research and Training Network ``The Physics of the Intergalactic
Medium''.
\end{acknowledgements}

\begin{figure*}
\includegraphics[angle=-90,width=17cm]{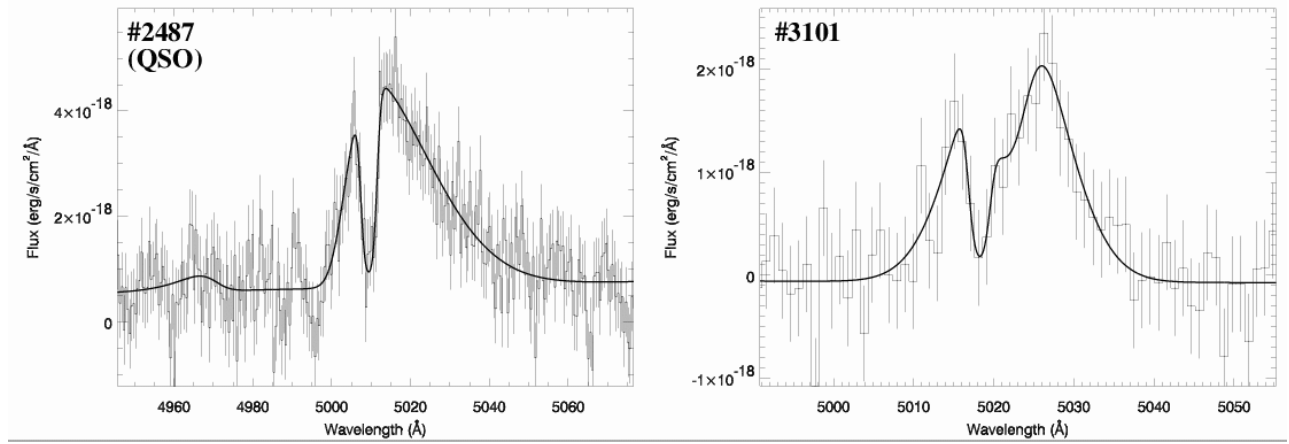}
\caption{\label{spec24873101} Spectra of emitter \#2487 (left) and \#3101 (right).}
\end{figure*}

\begin{figure*}
\includegraphics[angle=-90,width=17cm]{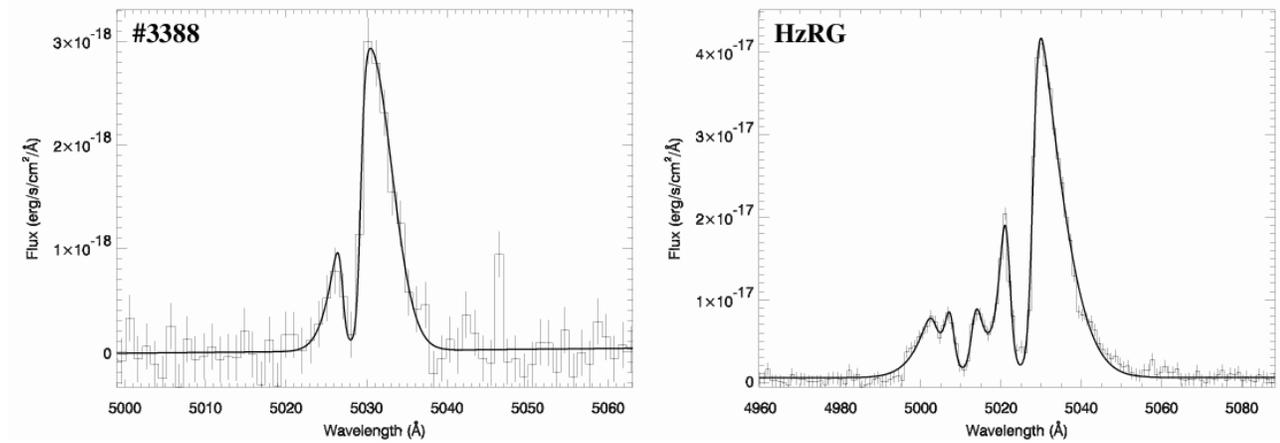}
\caption{\label{spec33881687} Spectra of emitter \#3388 (left) and HzRG (right).}
\end{figure*}

\end{document}